\documentclass[11pt,a4paper]{article}
\pdfoutput=1
\usepackage{jheppub}
\usepackage[utf8]{inputenc}
\usepackage{dcolumn}
\usepackage{bm,amsfonts,amsthm,amsmath,amssymb}
\usepackage{graphicx}
\usepackage{subfigure}

\newcommand{\ak}{a_{\vec{k},\lambda}}
\newcommand{\adk}{a^\dagger_{\vec{k},\lambda}}
\newcommand{\admk}{a^\dagger_{-\vec{k},\lambda}}

\title{A discrete series gauge field at the late-time boundary of $dS_4$}            
\author[a]{Manizheh Botshekananfard,}
\author[a]{Elif Büşra Güraksın,}
\author[b]{Vasileios A. Letsios,}
\author[a,c]{Gizem \c{S}eng\"or}
\affiliation[a]{Physics Department, Boğaziçi University\\
34342 Bebek, İstanbul Turkey}
\affiliation[c]{Feza Gürsey Center}
\affiliation[b]{Physique de l’Univers, Champs et Gravitation, Université de Mons – UMONS \\
Place du Parc 20, 7000 Mons, Belgium}
\emailAdd{manizheh.botshekananfard@bogazici.edu.tr}
\emailAdd{elifbguraksin@gmail.com}
\emailAdd{vasileios.letsios@umons.ac.be}
\emailAdd{gizem.sengor@bogazici.edu.tr}
\abstract{We study the free Maxwell field on the planar patch of four-dimensional de Sitter spacetime ($dS_4$). We review its bulk canonical quantization in the Bunch-Davies vacuum, and we give a representation-theoretic viewpoint by studying the transformation properties of single-particle states under infinitesimal dS transformations. By taking the late-time limit, we identify two late-time operators with scaling dimensions $\Delta=1$ (leading) and $\Delta=2$ (subleading).  We  introduce CFT-inspired  inner products invariant under the dS group ($SO(4,1)$) for states created by late-time operators acting on the Bunch-Davies vacuum. We explain how the unitary discrete series representations of $SO(4,1)$ associated with the photon are furnished by both operators, in contrast with the Maxwell field on $AdS$ where the $\Delta=1$ operator corresponds to a non-normalizable mode. We  also explain how the corresponding discrete series representations of $SO(4,1)$ split into a direct sum of representations corresponding to two helicities $+1$ and $-1$.
  This is achieved by taking advantage of the dS invariance of the self-dual and anti-self-dual sectors of the photon field strength. In our outlook, we draw inspiration from a recent proposal for the microscopic description of $dS_4$ higher-spin gravity where conformal gauge fields in 3 dimensions play a central role, and we investigate a possible connection of the  $\Delta=1$ late-time operator with  a conformal spin-1 gauge field in 3 dimensions.}
\begin{document}

\maketitle

\newpage

\begin{abstract} 

\end{abstract}

\section{Introduction}
\label{sec:Intro} 

Earlier on in \cite{unitarity} we started identifying late-time operators at the late-time boundary of de Sitter (dS) spacetime in a setting similar to the study of inflationary scalar perturbations: a scalar field introduced perturbatively and calculations carried on in the conformal Poincaré/Planar patch. Our main goal was to establish that these late-time operators correspond to unitary irreducible representations of the isometry group of dS spacetime. Our underlying motivation is that identifying such a list of late-time boundary operators and their correlation functions, as in \cite{twopoint}, might eventually provide input for constructing dS/CFT dualities, by giving a list of what type of operators and correlation functions to expect from the dual  conformal field theory (CFT) based on input from the de Sitter quantum field theory side.  Eventually, such lists of late-time operators and correlation functions may lead to practical studies of de Sitter Holography envisioned in \cite{Witten:2001kn,Strominger:2001pn,Maldacena:2002vr} and on understanding why the results of the inflationary observations are the way they are, as depicted in \cite{vanderSchaar:2003sz,Pajer:2016ieg}.  {For an up-to-date  review on the status of dS/CFT see, e.g., the introduction of \cite{Bzowski:2023nef}.} The most concrete realization of a dS/CFT-like duality, in the sense that both the bulk and the microscopic theories are independently defined, concerns the theory of higher-spin gravity on $dS_4$ \cite{Anninos:2017eib, Anninos:2011ui, Anninos:2026hia, Neiman:2014npa, David:2020ptn}.

By studying scalar fields with various values of the mass  on a fixed  de Sitter background in arbitrary dimensions, in \cite{unitarity} we introduced late-time operators corresponding to the \emph{principal} and \emph{complementary} series unitary irreducible representations of the de Sitter group $SO(d+1,1)$, i.e. the isometry group of $dS_{d+1}$, for heavy and light masses, respectively.%
\footnote{For related work concerning late-time operators in the principal and complementary series see \cite{Hogervorst:2021uvp, SalehiVaziri:2024joi}.} %
In particular, the identification of the unitary irreducible representations was made by finding the appropriate inner product that rendered states created by late-time operators normalizable.  
In \cite{twopoint} we extended this study to the case of exceptional series by studying the massless scalar%
\footnote{Here we refer to the  massless minimally coupled non-compact scalar field. The compact case in de Sitter spacetime has been studied in \cite{Letsios:2026ypo, Chakraborty:2023eoq, Chakraborty:2025mhh}.}~%
late-time operators on $dS_4$. 
  In \cite{Sengor:2023buj}, motivated by \cite{Anninos:2023lin}, we extended our list to discrete series late-time operators by considering the massless scalar on $dS_2$. Here we wish to enlarge our list further by focusing on the \emph{discrete series} of $SO(4,1)$. 

From the perspective of the method of induced representations, unitary irreducible representations arise from representations of invariant subgroups. For the group $SO(d+1,1)$ there are two different invariant subgroups. While principal and complementary series representations are induced from the subgroup consisting of dilatation, rotations and special conformal transformations, discrete series are induced from the maximally compact subgroup $SO(d+1)$. Two of the distinguishing features that set discrete series representations apart are that: they arise only when  the rank of the de Sitter group equals the rank of the maximally compact subgroup, which restricts to cases of even $d+1$, and they come in two types, which are not unitarily equivalent and they are the mirror image of each other \cite{Dobrev:1977qv}. In the present paper, we explain how the two inequivalent discrete series representations of $SO(4,1)$ (furnished by states of opposite helicity) manifest themselves at the late-time boundary of $dS_{4}$ in the case of the massless spin-1 field. We will often refer to these two discrete series as: \emph{discrete series $-$} and \emph{discrete series $+$}, corresponding to negative and positive helicities, respectively. Apart from principal, complementary and discrete series, there also exist exceptional series  representations of $SO(d+1,1)$. For example, totally symmetric (partially) massless tensor gauge fields on $dS_{d+1}$ correspond to the spinning exceptional series. For general $d+1 \geq  5$, exceptional series are different from discrete series (recall that the latter exist only when $d+1$ is even). However, for $d+1=4$, discrete series coincide with exceptional series. For more details see \cite{Hinterbichler:2026xqf, Dobrev:1977qv, Basile:2016aen, sun2021note, Penedones:2023uqc, Sengor:2022kji}.

The discrete series of $SO(4,1)$ are very interesting as they correspond to (massless and partially massless) spinning \emph{gauge fields} on four-dimensional dS spacetime. The study of gauge fields on dS spacetime is a topic of growing interest -- see: Refs. \cite{Silva_note, Loparco:2025azm, Garidi:2006ey, Lindsay:2026emf} for discussions concerning the massless spin-1 field (photon)\footnote{See \cite{Gazeau:1999xn} for discussions on massive spin-1 fields in the principal series.}, Ref. \cite{Higuchi:1991tn, Higuchi:1986py} for discussions on the massless spin-2 field (graviton), Ref. \cite{Anninos:2017eib} for massless higher-spin bosonic gauge fields, Refs. \cite{Higuchi:STSHS, Higuchi:1987hw, Basile:2016aen} for massless and partially massless higher-spin gauge fields, Refs. \cite{Letsios:2022tsq, Letsios:2023awz, Letsios:2023qzq, Anninos:2025mje, Schaub:2024rnl} for massless higher-spin fermionic gauge fields, and Refs. \cite{Higuchi-Letsios, Anninos:2025mje, Letsios:2023tuc, Deser:2003gw, Boulanger:2026wnw} for theories combining bosonic and fermionic gauge fields.\footnote{See also \cite{Schaub:2024rnl, sun2021note, Penedones:2023uqc} for nice representation-theoretic reviews.}
In the case of two-dimensional dS spacetime, where the relevant group is $SO(2,1)$, discrete series correspond to shift-symmetric scalars  (including the massless minimally coupled case, as well as the ``tachyonic'' scalars), where some of their modes are treated as pure-gauge \cite{Joung:2007je, Bros:2010wa, Higuchi:1987hw, Higuchi:STSHS, Letsios:2026ypo, Farnsworth:2024yeh}. Fermionic discrete series of $SL(2, \mathbb{R}) \cong Spin(2,1)$ are realized in terms of imaginary-mass spinors (including the zero-mass spinor field) on two-dimensional dS spacetime which also enjoy a 
fermionic shift symmetry \cite{Letsios:2025pqo}. Interestingly, such discrete series fermions  might prove useful towards constructing further solvable models in $dS_2$ \cite{Galati:2026btt}. Examples of interacting theories with discrete series fields in 2 dimensions have been given in \cite{Anninos:2023lin, Anninos:2023exn, Anninos:2024fty}.

  In the context of free QFT for  massless  spinning  gauge fields in global $dS_4$, it is clear how to identify the single-particle Hilbert space with discrete series irreducible representations of $SO(4,1)$ \cite{Letsios:2022tsq, Letsios:2023qzq, Letsios:2023awz, Higuchi-Letsios, Higuchi:1991tn, Higuchi:1987hw,Higuchi:2025pbc, Higuchi:STSHS}. In this setting, the group-theoretic properties of the mode solutions on global $dS_4$ play a central role, as one can rigorously show that the mode solutions themselves furnish unitary irreducible representations of  $SO(4,1)$ (this is true for massive, massless and partially massless cases). However, in the planar patch (\ref{conformal planar}), which is relevant to inflationary cosmology, it is less clear how the representations formed by bulk mode solutions can be identified with unitary irreducible representations of $SO(4,1)$. In fact, the planar patch covers only half of dS spacetime, so one expects that the mode solutions furnish unitary irreducible representations of the patch-preserving subgroup. However, these modes are still expected to form  representations of the dS algebra $so(4,1)$, although to the best of our knowledge, a detailed analysis is missing from the literature. Nevertheless, on the late-time boundary  of planar $dS_4$, the realization of unitary irreducible representations on the Hilbert space is more straightforward  to understand, where one makes  use of a CFT-inspired late-time inner product \cite{sun2021note, Loparco:2025azm, twopoint, Corfu2022proceed, unitarity}. Our present work focuses on the late-time realization of $SO(4,1)$ representations in the photon Hilbert space.

\paragraph{Results.} In the present paper, we consider the free Maxwell gauge potential (photon field) on $dS_4$. We fully fix the gauge at the classical level and we review the bulk canonical quantization in the Bunch-Davies vacuum on planar $dS_4$. We give a representation-theoretic viewpoint by studying the transformation properties of single-particle states under infinitesimal dS transformations in the standard bulk QFT Hilbert space.
  Then, we take the late-time limit and we explain how this gives rise to two late-time spin-1 operators with scaling dimensions $\Delta=1$  ($\sim$ boundary gauge field) and $\Delta=2$ ($\sim$ conserved current). We give explicit expressions of the late-time operators in terms of bulk creation and annihilation operators.  We also introduce $SO(4,1)$-invariant, CFT-inspired  inner products, following \cite{Dobrev:1977qv}, for `late-time states', i.e. states created by acting with late-time operators on the Bunch-Davies vacuum state. The Hilbert space spanned by late-time states, equipped with the CFT-inspired norm, carries representations of $SO(4,1)$. Using the corresponding intertwiner for the $\Delta=1$ operator to define an inner product, we find that  $\Delta=1$ late-time states have positive norm, furnishing the photon unitary discrete series representation of $SO(4,1)$.%
  \footnote{See also \cite{Loparco:2025azm, Silva_note, Lindsay:2025chz, Taylor:2026ase} for recent discussions on unitary irreducible representation of the dS group associated with the photon.}~
  This is in contrast with the Maxwell field on $AdS_4$ where the $\Delta=1$ operator corresponds to a non-normalizable mode. In the $\Delta=2$ case, we  take advantage of the known isomorphism \cite{Dobrev:1977qv} between the $SO(4,1)$ representation spaces of conserved currents and boundary gauge fields (with pure-gauge modes modded out), and we find another realization of the unitary discrete series representation associated with the photon on $dS_4$.  We  also explain how the $SO(4,1)$ representations realized on the bulk QFT Hilbert space, as well as on the space of late-time states, split into a direct sum of representations corresponding to two helicities $+1$ and $-1$.
  This is achieved by taking advantage of the dS invariance of the self-dual and anti-self-dual sectors of the photon field strength. The phenomenon of reducibility of massless representations, and their split into a direct sum of irreducible representations, which we point out in this paper, are understood in the case of global $dS_4$ for any strictly and partially massless gauge potential \cite{Letsios:2022tsq, Letsios:2023qzq, Letsios:2023awz, Higuchi-Letsios, Higuchi:1991tn, Higuchi:1987hw,Higuchi:2025pbc, Silva_note, Hinterbichler:2026xqf, Lindsay:2025chz, Taylor:2026ase}, but was missing in the case of planar $dS_4$. Last, we draw inspiration from the recently proposed microscopic ``gluing'' description of $dS_4$ higher-spin gravity in terms of the sphere path integrals \cite{Anninos:2026hia}, where conformal gauge fields in 3 dimensions play a central role, and we investigate a possible connection of the leading $\Delta=1$ late-time operator with the linearized fluctuations of  a conformal spin-1 gauge field in 3 dimensions.


$$ \textbf{Outline of the paper}  $$

\noindent In section \ref{sec:Abelian gague field}, we present the basic background material for the massless spin-1 gauge potential $A_\mu$ on the planar coordinate system (\ref{conformal planar}) of $dS_4$. We fix the gauge completely and we  decompose the bulk field into Bunch-Davies modes. We normalize the modes with respect to the Klein-Gordon norm, and perform canonical quantization of the gauge-fixed gauge potential. Earlier work on this is \cite{Cotaescu:2008hv}. Then, we focus on the late-time boundary of $dS_4$ and identify the two gauge-potential late-time operators $\alpha_j(\vec{k})$ and $\beta_j(\vec{k})$ with scaling dimensions 1 and 2, respectively. 
\\
\\
\noindent In section \ref{sec:Warm up: The QFT Hilbert space}, we start by reviewing the standard Hilbert space structure of the Maxwell gauge potential in the bulk of planar $dS_4$, and present a representation-theoretic viewpoint. We explain how quantum dS charges generate $so(4,1)$ transformations on the Hilbert space, where the standard bulk QFT norm is positive and $so(4,1)$-invariant. Then, we introduce briefly the notion of a CFT-inspired, $SO(4,1)$-invariant `late-time inner product' for the exceptional series, following \cite{Dobrev:1977qv}. Such inner products will be used to study representation-theoretic properties of states created by acting with the late-time operators on the Bunch-Davies vacuum in the following section.
\\
\\
\noindent In section \ref{sssec:The corresponding discrete series late-time operators}, we start by giving background material for exceptional series representations of $SO(d+1,1)$ -- exceptional series coincide with the discrete series for $d+1=4$, which is our case of interest. We then identify the exceptional series representation spaces of $SO(4,1)$ where our $\Delta \in \{1,2\}$ gauge-potential late-time operators live in. Then, we focus on the space of late-time states (states created by late-time operators acting on the Bunch-Davies vacuum), and we introduce CFT-inspired, $SO(4,1)$-invariant inner products following \cite{Dobrev:1977qv}. We identify the inner products with respect to which states created by $\alpha_j$ and $\beta_j$ have finite positive norm, giving rise to the unitary discrete series representation of $SO(4,1)$ associated with the photon on $dS_4$\footnote{See also \cite{Silva_note, Loparco:2025azm} for recent representation-theoretic discussions concerning the photon on dS spacetime.}. 
\\
\\
\noindent Section \ref{sec:field strength and anti-self-duality} focuses on the photon field strength and its (anti-)self-duality properties. We first give the bulk mode expansion of the field strength quantized in the Bunch-Davies vacuum. We then explain how the field strength can be split into self-dual and anti-self-dual parts, and we give the mode expansion for each part. Last, we show that the self-dual field strength does not mix with the anti-self-dual field strength under $so(4,1)$. 
\\
\\
\noindent  Section \ref{sec:Direct sum of representations in the Hilbert space} focuses on the split of the  space of states into a direct sum of representations furnished by states with opposite helicity ($+1$ and $-1$). Taking advantage of the previously established $so(4,1)$-invariance of the self-dual and the anti-self-dual sector of the field strength, we show that the standard bulk QFT Hilbert space splits into a direct sum of $so(4,1)$ representations furnished by states with opposite helicity. Then, we  take the late-time limit of the field strength. We identify the late-time operators of the self-dual and anti-self-dual field strengths, and we express them in terms of the late-time operators of the gauge potential. This allows us to split the gauge-potential late-time operators into two parts $\alpha_j(\vec{k})=\alpha_j^{(+)}(\vec{k})+\alpha_j^{(-)}(\vec{k})$ and $\beta_j(\vec{k})=\beta_j^{(+)}(\vec{k})+\beta_j^{(-)}(\vec{k})$. We identify $\alpha_j^{(+)}$ and $\beta_j^{(+)}$ as the gauge-potential late-time operators associated with the anti-self-dual field strength. Similarly, we identify $\alpha_j^{(-)}$ and $\beta_j^{(-)}$ as the gauge-potential late-time operators associated with the self-dual field strength. Using again the $so(4,1)$-invariance of the self-dual and anti-self-dual sectors, we show that states created by the late-time operator $\alpha^{(-)}_j$ do not mix with states created by the late-time operator $\alpha^{(+)}_j$, and similarly states created by the late-time operator $\beta^{(-)}_j$ do not mix with states created by $\beta^{(+)}_j$ under dS transformations.
\\
\\
\noindent  In section \ref{sec:Conclusions and Outlook}, we summarize our results and we give the commutation relations of the late-time operators. Then we discuss the possible connection of the leading $\Delta=1$ operator with conformal gauge fields in 3 dimensions (inspired by recent findings concerning dS higher-spin gravity \cite{Anninos:2026hia}), and we also discuss future directions.


\section{Abelian gauge field}\label{sec:Abelian gague field}
We will work in the conformal planar patch, where the line element is 
\begin{align} \label{conformal planar}ds^2=\frac{-d\eta^2+d\vec{x}^2}{H^2|\eta|^2},~~\eta\in(-\infty,0).\end{align}
Our conventions are such that $\hbar=c=1$, which sets the dimension of $H$ to be that of $mass$ and spacetime coordinates have dimension $mass^{-1}$. Here length has been identified with $mass^{-1}$. As such the metric components $g_{\mu\nu}$ are dimensionless. Greek tensor indices $\mu,\nu,...$ take  values in $\{\eta,x ,y, z\}$, while Latin indices are purely spatial.

Let us consider a free real Abelian $U(1)$ gauge field with the following action on a $dS_4$ background
\begin{align}
    S_A=-\frac{1}{4}\int d^4x\sqrt{-g}F_{\mu\nu}F^{\mu\nu},~~~F_{\mu\nu}=\partial_\mu A_\nu-\partial_\nu A_\mu.
\end{align}
 The equations of motion have the form
\begin{align}
    \frac{1}{\sqrt{-g}}\partial_\mu\left(\sqrt{-g} F^{\mu\nu}\right)=0.
\end{align}
Both the action and the field equations are invariant under $U(1)$ gauge transformations
\begin{align}
    \delta A_{\mu}=\partial_{\mu}\lambda,
\end{align}
where $\lambda$ is a smooth scalar gauge function.

The first thing to note is that because of the antisymmetric nature of $F_{\mu\nu}$ no $\partial_\eta A_\eta$ terms appear in the above action. So one expects to find a constraint which, in the covariant form, appears as 
\begin{align}
    \frac{1}{\sqrt{-g}}\partial_\mu\left(\sqrt{-g}F^{\mu \eta}\right)=0.
\end{align}
More specifically, $A_\eta$ generates the following constraint\footnote{We raise and lower indices for the derivatives as $\partial^{i} = g^{ij} \partial_{j} = H^2|\eta|^{2} \delta^{ij} \partial_{j} $. Similarly, $A^{i} = g^{ij} A_{j} = H^2|\eta|^{2} \delta^{ij} A_{j}$. In our use of the dot and vector notation on derivatives we employ flat spacetime conventions. That is $\dot{}\equiv \partial_\eta$ and $\vec{\partial}^2\equiv\partial^2_x+\partial^2_y+\partial^2_z$.}
\begin{align}\label{constraint}
    -\partial_i\partial^iA_\eta+\partial_\eta\left(\partial_i A^i\right)=0.
\end{align}
This is a secondary constraint obtained by demanding that the primary constraint $\pi_\eta\equiv \frac{\partial \mathcal{L}}{\partial\left( \partial_\eta{A_\eta}\right)}=0$ is preserved in time, that is $\partial_\eta\pi_\eta=0$.

Let us pick the gauge  $A_\eta=0$. The constraint \eqref{constraint} demands that for this gauge choice we should also set $\partial_iA^ i=0$. Note that we also have $g^{jm}\nabla_{j} A_{m}=0$ since the purely spatial Christoffel symbols $\Gamma^{\ell}_{jm}$ are zero. Once this gauge has been picked, there is no gauge freedom left.
In the coordinate system \eqref{conformal planar}, with this gauge choice, the equation of motion is
\begin{align}\label{eq:AUieom}\ddot A^i+\frac{4}{|\eta|}\dot A^i-\vec{\partial}^2A^i~{+\frac{6}{\eta^2}A^i}=0,\end{align}
where the dot denotes partial derivative with respect to conformal time $\eta$, and $\vec{\partial}^2 \equiv \partial_x^2 + \partial_y^2 + \partial_z^2$. Although the components with the upper index are ``aware'' of the curvature, the components with lower index obey the same wave equation as in flat space
\begin{align} \label{eom:ALi xspace}
\ddot A_i-\vec{\partial}^2 A_i=0.\end{align}
Equivalently, we can re-express the on-shell conditions satisfied by the field as
\begin{align}
   &\Box A_{\mu}= 3 H^2\,A_\mu , \label{covariant field EOM}\\
   &A_{\eta}=g^{ij}\nabla_i A_j=0, \label{gauge condtions under covariant EOM}
\end{align}
where $\Box =  g^{\mu \nu}\nabla_{\mu}\nabla_{\nu}$.

For practical purposes we will be interested in the quantized field, decomposed into Fourier modes in terms of polarization vectors, $\hat{\epsilon}^{(\lambda)}_i(\vec{k} )$, and annihilation and creation operators, $\ak,\adk$, with helicity label $\lambda=\pm$. 

\subsection{The classical $A_i$ modes}
\label{subsec:ALi modes}

We decompose the vector-field components $A_i$ as
\begin{equation}  \label{eq:ALi modedecomp} A_i(\vec{x},\eta)=\sum_{\lambda=\pm}\int \frac{d^3k}{(2 \pi)^{3/2}}\left[\mathcal{A}_{\vec{k},\lambda} A_i^{(\lambda,\vec{k})}+\mathcal{A}^*_{\vec{k},\lambda} \left(A_i^{(\lambda,\vec{k})}\right)^*\right],
\end{equation}
where, at the classical level, the $\mathcal{A}_{\vec{k},\lambda}$'s are complex coefficients with dimensions of $mass^{-3/2}$,  labeled by  momentum $\vec{k}$ and polarization $\lambda$ of each mode%
\footnote{In units where $\hbar=c=1$ the action should be dimensionless and $H$ has dimension $mass$. Accordingly, the vector field components $A_i$ have dimension $mass$. While $d^3k$ has dimension $mass^3$, the complex constants $\mathcal{A}_{\vec{k},\lambda}$ have dimension $mass^{-3/2}$. When we canonically quantize the field below, we require that the operators $a_{\vec{k},\lambda}$, $a^\dagger_{\vec{k},\lambda}$ have dimensions $mass^{-\frac{3}{2}}$. The modes, $A_j^{(\lambda,\vec{k})}$, have dimension $mass^{-\frac{1}{2}}$, stemming from the factor of $|\eta|^{1/2}$ in (\ref{eqn:modes}). The delta function $ \delta^{(3)}(\vec{k}'-\vec{k})$ has dimensions off $mass^{-3}$, while $ \delta^{(3)}(\vec{x}'-\vec{x})$ has dimensions off $mass^{3}$.}. %
The physical (i.e. non-pure-gauge) positive frequency modes\footnote{The Hankel function that arises in this particular case is conveniently expressed in terms of an exponential
\begin{align}
    H_{\frac{1}{2}}^{(1)}(k |\eta|)=-i\sqrt{\frac{2}{\pi}}\frac{e^{ik|\eta|}}{\sqrt{k|\eta|}}.
\end{align}} are \cite{Cotaescu:2008hv}
\begin{align}\label{eqn:modes}
A_{j}^{(\lambda,\vec{k})}(\vec{x},\eta) = \frac{\mathcal{C}}{(2 \pi)^{3/2}}\, |\eta|^{1/2} H_{\frac{1}{2}}^{(1)}(k |\eta|)~ \hat{\epsilon}_{j}^{(\lambda)}(\vec{k}) ~e^{i\vec{k}\cdot\vec{x}},
\end{align}
and they obey Bunch-Davies initial conditions, i.e. they decay as $\eta\to -\infty(1+i\epsilon)$. Accordingly, ${A_{j}^{(\lambda,\vec{k})}}^*(\vec{x},\eta)$ correspond to the negative frequency modes. The normalization factor $\mathcal{C}$ is dimensionless, which we will fix below by Klein-Gordon normalization. Similar decompositions for a vector field on FRLW backgrounds can be found in the cosmology \cite{Lee:2016vti,Anber_2010,Armendariz_Picon_2008} and in dS literature  \cite{Higuchi:1986py,Saharian:2013yya}. We choose to write the solution in terms of Hankel functions for later convenience. This form also makes it easier to compare with a more general treatment in arbitrary dimensions \cite{Lee:2016vti}. Solution \eqref{eqn:modes} also exhibits the expected scaling behavior in the late-time limit, as we will explore in section \ref{sssec:Late-time scaling behaviour}.

The label $\lambda = \pm$ corresponds to the two helicities $\pm 1$. The polarization vectors ${\epsilon}^{(\lambda)}_j(\vec{k})$ obey 
\begin{align}
    \label{eq:pol vec} &\hat{\epsilon}^{(\pm)}_j(\vec{k})^{*}=\hat{\epsilon}^{(\mp)}_j(\vec{k}),~~k_i\hat{\epsilon}^{(\pm)}_i(\vec{k})=0,~~\epsilon_{lij}~k_i~\hat{\epsilon}^{(\pm)}_j(\vec{k})=\mp i ~k~\hat{\epsilon}^{(\pm)}_l(\vec{k})
\end{align}
where $\epsilon_{ijk}$ is the Levi-Civita symbol (repeated indices are summed over). They are normalized as
\begin{align}
\label{eqn:pol normalization}    \delta^{ij} \hat{\epsilon}^{(\lambda)}_{i}(\vec{k})^{*}~\hat{\epsilon}^{(\lambda')}_{j}(\vec{k})=\delta_{\lambda \lambda'},
\end{align}
and they satisfy the completeness relation
\begin{align}\label{completeness of polarisation vecs}
    \sum_{\lambda=\pm} \hat{\epsilon}^{(\lambda)}_{i}(\vec{k}) \hat{\epsilon}^{(\lambda)}_{j}(\vec{k})^{*} = \delta_{ij}-\frac{k_{i}k_{j}}{k^{2}}.
\end{align}
Moreover, polarization vectors satisfy
\begin{align}\label{eqn:epsilon-k}
   \hat{\epsilon}_j^{(\lambda)}(-\vec{k})=\hat{\epsilon}_j^{(-\lambda)}(\vec{k}),
\end{align}
as well as%
\footnote{One can confirm the property \eqref{eqn:epsilon cross prop} as follows. Contracting the completeness relation (\ref{completeness of polarisation vecs}) with $\frac{k_{m}}{k}\epsilon_{\ell}\,^{mj}$ and using $\frac{1}{k}\epsilon_{\ell}\,^{mj}~k_m~\hat{\epsilon}^{(-\lambda)}_j(\vec{k})=+ i \lambda ~\hat{\epsilon}^{(-\lambda)}_\ell(\vec{k})$  \eqref{eq:pol vec}, one finds
\begin{align}\label{eqn:epsilon prop}
  \sum_{\lambda = \pm} i~ \lambda  ~ \hat{\epsilon}^{(\lambda)}_{i}(\vec{k}) \hat{\epsilon}^{(-\lambda)}_{\ell} (\vec{k})= i~ \left(  ~ \hat{\epsilon}^{(+)}_{i}(\vec{k}) \hat{\epsilon}^{(-)}_{\ell} (\vec{k}) - \hat{\epsilon}^{(-)}_{i}(\vec{k}) \hat{\epsilon}^{(+)}_{\ell} (\vec{k})\right)=\epsilon_{i\ell m} \frac{k^{m}}{k}.
\end{align}
Contracting equation \eqref{eqn:epsilon prop} with $\epsilon^{ai \ell}$, and readjusting the terms, we find
\begin{align*}
 \epsilon^{ai \ell}  ~ \hat{\epsilon}^{(+)}_{i}(\vec{k}) \hat{\epsilon}^{(-)}_{\ell} (\vec{k})=\left( \hat{\epsilon}^{(+)}(\vec{k}) \times  \hat{\epsilon}^{(-)}\right)^{a}=-\left( \hat{\epsilon}^{(-)}(\vec{k}) \times  \hat{\epsilon}^{(+)}\right)^{a} =-i \frac{k^{a}}{k}
\end{align*}
in agreement with \eqref{eqn:epsilon cross prop}.}, 
\begin{equation}
\label{eqn:epsilon cross prop}
\hat{\epsilon}^{(-)}(\vec{k})\times\hat{\epsilon}^{(+)}(\vec{k})=i\hat{k} ,
\end{equation}
where  $\hat{k} = \vec{k}/k$.
%
The components of the polarization vectors are given by
\begin{align}
  &\hat{\epsilon}^{(\lambda)}_{1}(\vec{k}) =c~(k-k_1)(k+k_1) ,\nonumber\\
&\hat{\epsilon}^{(\lambda)}_{2}(\vec{k}) =-c~(k_1 \, k_{2}-i \lambda \,k\,k_3), \nonumber\\
\label{eqn:epsilon comps}&\hat{\epsilon}^{(\lambda)}_{3}(\vec{k}) =-c~(k_1 \, k_{3}+i \lambda \,k\,k_2) 
\end{align}
where the normalization factor $c$ is 
\begin{align}
    c= \frac{1}{k~\sqrt{2(k-k_1)(k+k_1)}}.
\end{align}
From these expressions, it becomes clear that $\hat{\epsilon}^{(-\lambda)}_{j}(\vec{k})=\hat{\epsilon}^{(\lambda)}_{j}(\vec{k})^* = \hat{\epsilon}^{(\lambda)}_{j}(-\vec{k})$. With the conventions of \eqref{eqn:epsilon comps}, for a mode propagating with $\vec{k}=(0,0,k)$, that is purely in the third direction, one has $\hat{\epsilon}^{(\lambda)}_{1}(\vec{k})=\frac{1}{\sqrt{2}}$, $\hat{\epsilon}^{(\lambda)}_{2}(\vec{k})=\frac{i}{\sqrt{2}}\lambda$ and $\hat{\epsilon}^{(\lambda)}_{3}(\vec{k})=0$.

\paragraph{Klein-Gordon inner product and normalization.} Let us normalize the physical modes using the Klein-Gordon inner product 
\begin{align}
  \label{eqn:KGinner prod}  \left(A^{(\lambda,\vec{k})}, A^{(\lambda',\vec{k}')} \right)_{KG}&=-i \int {d^{3}x~\sqrt{-g}} g^{im}\left( A^{(\lambda,\vec{k})*}_{i} \nabla^{\eta}A^{(\lambda',\vec{k}')}_{m} -\nabla^{\eta}A^{(\lambda,\vec{k})*}_{i} A^{(\lambda',\vec{k}')}_{m} \right) \nonumber\\
    &=i \int {d^{3}x} ~\delta^{im}\left( A^{(\lambda,\vec{k})*}_{i} \partial_{\eta}A^{(\lambda',\vec{k}')}_{m} -\partial_{\eta}A^{(\lambda,\vec{k})*}_{i} A^{(\lambda',\vec{k}')}_{m} \right).
\end{align}
The Klein-Gordon inner product is time-independent because it corresponds to the (Noether) charge associated with the conserved Klein-Gordon current
\begin{align}
  \label{eqn:KG current}  J^{\mu}_{KG}\left(A^{(\lambda,\vec{k})}, A^{(\lambda',\vec{k}')} \right)&=-i~g^{\kappa \rho}\left( A^{(\lambda,\vec{k})*}_{\kappa} \nabla^{\mu}A^{(\lambda',\vec{k}')}_{\rho} -\nabla^{\mu}A^{(\lambda,\vec{k})*}_{\kappa} A^{(\lambda',\vec{k}')}_{\rho} \right) ,~~~\nabla_\mu J^{\mu}_{KG}=0.
\end{align}
The Klein-Gordon inner product is then expressed as
\begin{align}\label{KG product in terms of KG current}
 \left(A^{(\lambda,\vec{k})}, A^{(\lambda',\vec{k}')} \right)_{KG} = \int d^3x \sqrt{-g}~ J^{\eta}_{KG}\left(A^{(\lambda,\vec{k})}, A^{(\lambda',\vec{k}')} \right).
\end{align}
Note that the current $J^{\mu}_{KG}\left(A^{(1)}, A^{(2)} \right)$ is covariantly conserved for any two solutions $A^{(1)}_{\mu}, A^{(2)}_{\nu}$ in the Lorenz gauge ($\nabla^\mu A^{(1)}_\mu = \nabla^\mu A^{(2)}_\mu =0$), which includes the Coulomb gauge (\ref{gauge condtions under covariant EOM}) as a special case.

Substituting the explicit expressions of the modes into the inner product (\ref{eqn:KGinner prod}), and using
\begin{align} \label{identity for Hankels}
    \left(H^{(1)}_{\nu}(x)\right)^{*} \frac{d}{dx}H^{(1)}_{\nu}(x) - 
\left(\frac{d}{dx}H^{(1)}_{\nu}(x)\right)^{*} H^{(1)}_{\nu}(x) = \frac{4i e^{\pi ~Im(\nu)}}{\pi x}
\end{align}
as well as,
\begin{equation}
    \delta^{(3)}(\vec{k}'-\vec{k})=\int \frac{d^3x}{(2\pi)^3}e^{i(\vec{k}'-\vec{k})\cdot\vec{x}},
\end{equation}
we find 
   \begin{align}
    \left(A^{(\lambda,\vec{k})}, A^{(\lambda',\vec{k}')} \right)_{KG}&= |\mathcal{C}|^{2} \frac{4}{\pi} \delta^{(3)}(\vec{k}-\vec{k}')~\delta_{\lambda \lambda'}.
\end{align} 
We similarly find
\begin{align}
    \left(A^{(\lambda,\vec{k})*}, A^{(\lambda',\vec{k}')*} \right)_{KG}=- |\mathcal{C}|^{2} \frac{4}{\pi} \delta^{(3)}(\vec{k}-\vec{k}')~\delta_{\lambda \lambda'},~~ \left(A^{(\lambda,\vec{k})*}, A^{(\lambda',\vec{k}')} \right)_{KG}=0.
\end{align} 
We choose the normalization factor 
\begin{align}\label{KGnormalisation}
    \mathcal{C} = \frac{\sqrt{\pi}}{2},
\end{align}
so that positive-frequency modes are normalized to 1.


\paragraph{Completeness relation.} Using (\ref{completeness of polarisation vecs}) and (\ref{identity for Hankels}), it is easy to prove that the modes satisfy the following completeness relation with respect to the Klein-Gordon inner product  
\begin{align} \label{completeness reltn modes}
    \int {d^{3}k}\sum_{\lambda = \pm} \left(  A^{(\lambda,\vec{k})}_{j}(\vec{x},\eta) ~~  \partial_{\eta}A^{(\lambda,\vec{k})}_{m}(\vec{y},\eta)^{*} -(c.c.) \right) =&i \int \frac{d^{3}k}{(2 \pi)^{3}} \left(\delta_{jm} -\frac{k_{j}  k_{m}}{k^{2}}\right)~ e^{i \vec{k} \cdot (\vec{x} - \vec{y})}\\
    =&i  \left(  \delta_{jm} - \frac{1}{\delta^{np}\partial_{n}\partial_{p}}~~
    {\frac{\partial}{\partial{x^{j}}} \frac{\partial}{\partial{x^{m}}}  }    \right)\delta^{(3)}(\vec{x}   - \vec{y}),
\end{align}
where  $\partial_{n} = \partial / \partial x^n$. Note that $ \partial_{\eta} A^{(\lambda,\vec{k})}_{j}(\vec{x},\eta) = -i~k~ A^{(\lambda,\vec{k})}_{j}(\vec{x},\eta)$.

\subsection{Bulk quantization}
\label{sssec:bulk quantization}

We achieve bulk quantization by promoting the complex constants $\mathcal{A}_{\vec{k},\lambda}$ and their complex conjugates, $\mathcal{A}^*_{\vec{k},\lambda}$,  to  annihilation and creation operators,  $a_{\vec{k},\lambda}$ and $a^\dagger_{\vec{k},\lambda}$, respectively,   in equation \eqref{eq:ALi modedecomp}. Annihilation operators annihilate the (Bunch-Davies) vacuum $|0\rangle$ for all $\vec{k}$ and $\lambda$, while  creation operators create single-particle states with  certain momentum and polarization, and they satisfy canonical commutation relations, as:
\begin{subequations}\label{eqn:a ad intro}
\begin{align}
\ak|0\rangle=0,&~~~~\adk|0\rangle=|\lambda,\vec{k}\rangle,\\
\left[a_{\vec{k},\lambda},a^\dagger_{\vec{k}',\lambda'}\right]&=(2\pi)^3~\delta_{\lambda\lambda'}~\delta^{(3)}\left(\vec{k}-\vec{k}'\right).
\end{align}
\end{subequations}
The quantized field operator is expanded in modes as
\begin{align}\label{eqn:quantizedA}
    A_{j}(\vec{x},\eta) = \sum_{\lambda=\pm}\int \frac{d^{3}k}{(2   \pi)^{3/2}} \left( a_{\vec{k},\lambda} ~A^{(\lambda,\vec{k})}_{j}(\vec{x},\eta)+ a^{\dagger}_{\vec{k},\lambda} ~A^{(\lambda,\vec{k})}_{j}(\vec{x},\eta)^{*} \right).
\end{align}
The annihilation (and similarly the creation) operator is expressed using the inner product (\ref{eqn:KGinner prod}) as 
\begin{align}\label{annihilation op using KG}
    a_{\vec{k}, \lambda} = (2 \pi)^{3/2}~\left(   A^{(\lambda, \vec{k})}, A\right)_{KG}.
\end{align}

 By fixing our chosen gauge in the action functional, we can determine the conjugate momentum operator $\pi^{j}(\vec{x},\eta)$ as
\begin{align}
     \pi^{j}= \frac{\partial   \mathcal{L}}{\partial( \partial_{\eta}A_{j})} =-\sqrt{-g} g^{\eta \eta}~g^{mj} \partial_{\eta}A_{m}=\delta^{mj} \partial_{\eta}A_{m}. 
\end{align}
The equal-time commutation relations are 
\begin{align}
\left[A_i(\vec{x},\eta),{\pi}^j(\vec{y},\eta)\right]=i \left(  g_{i}\,^{j} - ~
    {\frac{\nabla_{i} \nabla^{j}}{\nabla^{n}\nabla_{n}}  }    \right)\delta^{(3)}(\vec{x}   - \vec{y}),
\end{align}
where $\nabla^{j}=g^{jm} \nabla_{m}$ and $\nabla^n \nabla_n = g^{nl} \nabla_n \nabla_l$. The commutation relations can be verified by expanding the fields in modes and using the completeness relation (\ref{completeness reltn modes}).
Note that, in the equal-time commutator above, $\nabla_{i} \nabla^{j} =\nabla_{i} \partial^{j}$ can be replaced by $ \partial_{i} \partial^{j}=g^{jm}\frac{\partial}{\partial x^i} \frac{\partial}{\partial x^m}$, since the Christoffel symbols with purely spatial components vanish.


\subsection{Introducing massless spin-1 late-time operators}
\label{sssec:Late-time scaling behaviour}
As one of the conformal boundaries of dS spacetime is located on the late-time slice, the late-time limit gives us information about what to expect from  the conformal field theory perspective. For symmetric, transverse-traceless fields $\phi_{i_1i_2...i_s}$ with integer spin $s$ on dS spacetime, the late-time limit is  \cite{baumann2018tasi, Arkani-Hamed:2015bza}
\begin{align}\label{eqn:latetime dic}
\lim_{\eta\to 0}\phi_{i_1i_2...i_s}(\vec{x}, \eta)=|\eta|^{\Delta_--s}\alpha_{i_1i_2...i_s}(\vec{x})+|\eta|^{\Delta_+-s}\beta_{i_1i_2...i_s}(\vec{x}),
\end{align}
where the $d$-dimensional late-time operators $\alpha_{i_1i_2...i_s}(\vec{x})$  and $\beta_{i_1i_2...i_s}(\vec{x})$ transform as conformal primaries with scaling dimensions $\Delta_-$ and $\Delta_+ = d- \Delta_-$, respectively, under the conformal group of the late-time boundary, $SO(d+1,1)$.
The limit is taken with respect to the time-like coordinate since dS boundaries are the early-time and late-time boundaries, reached along time-like directions. We are interested in the late-time boundary.

For representations of the de Sitter group $SO(d+1,1)$, as we will discuss in more detail in section \ref{sssec:The corresponding discrete series late-time operators}, the scaling dimension takes the following form
\begin{align}
    \Delta=\frac{d}{2}+c,
\end{align} 
where $c$ encodes information about mass and spin.
 For the case of the Maxwell field on $dS_4$, the late-time limit \eqref{eqn:latetime dic} reads
\begin{align}
    \label{eqn:latetime dic s1}\lim_{\eta\to 0}A_{i}(\eta,\vec{x})=|\eta|^0\alpha_{i}(\vec{x})+|\eta|^1\beta_{i}(\vec{x}),
\end{align}
where the scaling dimensions are
\begin{align}
\Delta_\alpha\equiv\Delta_-=1,~~\Delta_\beta\equiv\Delta_+=2.
\end{align}

As discussed earlier, with Bunch-Davies initial conditions, the  quantized $U(1)$ gauge field is expanded as
\begin{align}
    A_{j}(\vec{x},\eta) = \sum_{\lambda=\pm}\int \frac{d^{3}k}{(2   \pi)^{3/2}} \left( a_{\vec{k},\lambda} ~A^{(\lambda,\vec{k})}_{j}(\vec{x},\eta)+ a^{\dagger}_{\vec{k},\lambda} ~A^{(\lambda,\vec{k})}_{j}(\vec{x},\eta)^{*} \right),
\end{align}
with mode functions given by  
\begin{align}
A_{j}^{(\lambda,\vec{k})}(\vec{x},\eta) = \frac{1}{(2 \pi)^{3/2}}\frac{\sqrt{\pi}}{2} |\eta|^{1/2} H_{\frac{1}{2}}^{(1)}(k |\eta|)~ \hat{\epsilon}_{j}^{(\lambda)}(\vec{k}) ~e^{i\vec{k}\cdot\vec{x}}.
\end{align}
Equivalently, 
\begin{align}
\label{eq:ALi bulk}
A_j(\vec{x},\eta)=\frac{\sqrt{\pi}}{2}\sum_{\lambda=\pm}&\int \frac{d^3k}{(2 \pi)^{3}}\Bigg[\hat{\epsilon}_{j}^{(\lambda)}(\vec{k}) |\eta|^{1/2}H^{(1)}_\frac{1}{2}(k|\eta|)\ak +\hat{\epsilon}_{j}^{(\lambda)*}(-\vec{k}) |\eta|^{1/2}H^{(2)}_\frac{1}{2}(k|\eta|)\admk \Bigg]e^{i\vec{k}\cdot\vec{x}}.
\end{align}
Now let us determine the two late-time operators ${\alpha}_{j}(\vec{x})$ and ${\beta}_{j}(\vec{x})$, with scaling dimensions $\Delta_\alpha=1$ and $\Delta_\beta=2$, respectively, which appear in \eqref{eqn:latetime dic s1}. This is more easily seen in momentum space, where 
\begin{align}
\nonumber        \lim_{|\eta|\to0} A_j(\vec{x},\eta)=\int\frac{d^3k}{(2\pi)^{3}}e^{i\vec{k}\cdot\vec{x}}&\Bigg[|\eta|^0\left(i\,{{\frac{1}{\sqrt{2}}}}\sum_{\lambda=\pm}\left[\hat{\epsilon}_{j}^{(\lambda)*}(-\vec{k})\admk-\hat{\epsilon}_{j}^{(\lambda)}(\vec{k})\ak\right]k^{-\frac{1}{2}}\right)\\
\label{eqn:Alower latetime_v2}&+|\eta|\left({{\frac{1}{\sqrt{2}}}}\sum_{\lambda=\pm}\left[\hat{\epsilon}_{j}^{(\lambda)*}(-\vec{k})\admk+\hat{\epsilon}_{j}^{(\lambda)}(\vec{k})\ak\right]k^{\frac{1}{2}}\right)\Bigg].
\end{align}
%
We used that the late-time limit for the Hankel function  is \cite{NIST:DLMF,twopoint}
\begin{align}
\label{eqn:H1latetime}
\lim_{|\eta|\to0}H^{(1)}_{\frac{1}{2}}(k|\eta|)=\frac{1}{\Gamma(3/2)}\left(\frac{k|\eta|}{2}\right)^{\frac{1}{2}}-\frac{i\Gamma(1/2)}{\pi}\left(\frac{k|\eta|}{2}\right)^{-\frac{1}{2}}.
\end{align}
Comparing \eqref{eqn:Alower latetime_v2} with  \eqref{eqn:latetime dic s1}, we read off the momentum space late-time operators
\begin{subequations}
\label{eqn:introducing alpha beta k}
    \begin{align}
\nonumber\alpha_j(\vec{k})&=i{{\frac{1}{\sqrt{2}}}}\sum_{\lambda=\pm}\left[\hat{\epsilon}_{j}^{(\lambda)}(-\vec{k})^{*}~\admk-\hat{\epsilon}_{j}^{(\lambda)}(\vec{k})~\ak\right]k^{-\frac{1}{2}}\\
\label{eqn:introducingalphak}&=i{{\frac{1}{\sqrt{2}}}}\sum_{\lambda=\pm}\hat{\epsilon}_{j}^{(\lambda)}(\vec{k})\left[\admk-\ak\right]k^{-\frac{1}{2}},\\
\nonumber\beta_j(\vec{k})&={{\frac{1}{\sqrt{2}}}}\sum_{\lambda=\pm}\left[\hat{\epsilon}_{j}^{(\lambda)}(-\vec{k})^{*}~\admk+\hat{\epsilon}_{j}^{(\lambda)}(\vec{k})\ak\right]k^{\frac{1}{2}}\\
\label{eqn:introducingbetak}&={{\frac{1}{\sqrt{2}}}}\sum_{\lambda=\pm}\hat{\epsilon}_{j}^{(\lambda)}(\vec{k})\left[\admk+\ak\right]k^{\frac{1}{2}}.
    \end{align}
\end{subequations}
\label{eqn:intro alpha k-sp}
 From these, the position space late-time operators are obtained by the following Fourier transforms 
\begin{align}
\label{eqn:intro alpha beta xspace}\alpha_j(\vec{x})&=\int\frac{d^3k}{(2\pi)^3}e^{i\vec{k}\cdot\vec{x}}\alpha_j(\vec{k}),~~\beta_j(\vec{x})=\int\frac{d^3k}{(2\pi)^3}e^{i\vec{k}\cdot\vec{x}}\beta_j(\vec{k}).
    \end{align}
    Note that the late-time operators we introduced in \eqref{eqn:introducing alpha beta k} involve a sum over helicities, which suggests that they may be split further into operators with fixed helicity labels. We will elaborate on this observation further in section \ref{sec:Direct sum of representations in the Hilbert space}, which will lead us to split each operator in \eqref{eqn:introducing alpha beta k} into two distinct operators, as
     \begin{align}
    \alpha_j(\vec{k})&=\alpha_j^{(+)}(\vec{k})+\alpha_j^{(-)}(\vec{k}),\\
    \beta_j(\vec{k})&=\beta_j^{(+)}(\vec{k})+\beta_j^{(-)}(\vec{k}),
\end{align}
    corresponding to the direct sum of  irreducible representations with opposite helicity.

\subsection{Scaling dimensions of the late-time operators}
\label{subsec:Scaling dimensions of the late-time operators}
Lastly, let us check that the operators in \eqref{eqn:intro alpha beta xspace} indeed scale as promised. For a quick check, under a dilatation
\begin{align}
    x\to sx,~~\vec{k}\to s^{-1}\vec{k},
\end{align}
in accordance with their commutation relations \eqref{eqn:a ad intro} the annihilation and creation operators scale as
\begin{align}
    \label{eqn:a ad scaling} a_{\vec{k}/s,\lambda}=s^{3/2}\ak,~~a^\dagger_{\vec{k}/s,\lambda}=s^{3/2}\adk,
\end{align}
while the polarization vectors $\hat{\epsilon}_{j}^{(\lambda)}(\vec{k})$ being unit vectors do not scale.
Then, following the arguments of \cite{unitarity} and keeping in mind that $\delta^{(3)}\left([\vec{k}-\vec{k}']/s\right) = s^3 ~\delta^{(3)}\left(\vec{k}-\vec{k}'\right)$, one can check that indeed we have $\Delta_\alpha=1$ and $\Delta_\beta=2$ as promised. 

For a more thorough check of the scaling dimension, let us take a step back, and consider how the $so(4,1)$ algebra acts on late-time operators. The Killing vectors of $dS_4$ give a realization of the generators of the dS algebra. We identify Killing vectors with $so(4,1)$ generators following the conventions of \cite{particles}, as
\begin{subequations}\label{Killing vectors}
    \begin{align}
      \text{Dilatation:}~~D&=-\eta\partial_\eta-x^i\partial_i=-\overset{{\rm (D)}}{\xi}\\
      \text{Translations:}~~T_i&=-\partial_i=-\overset{{\rm (i-Trans)}}{\xi}\\
      \text{SCTs:}~~C_i&=\left(|\eta|^2-|\vec{x}|^2\right)\partial_i+2x^i\left(\eta \partial_\eta+x^l\partial_l\right)=-\overset{{\rm (i-SCT)}}{\xi}\\
      \text{Rotations:}~~M_{ij}&=-\left(x^i\partial_j-x^j\partial_i\right)=\overset{{\rm (rot-i\times j )}}{\xi},
    \end{align}
\end{subequations}
where `SCTs' stands for special conformal transformations.
Here, the indices of the generators should be understood as labels rather than spacetime indices.
To be precise, the right-hand sides of these equations concern the action of the generators on scalar fields on $dS_{4}$. When these generators act on spinning fields on $dS_{4}$, they are realized in terms of Lie derivatives with respect to the corresponding Killing vector. For example, for a vector field we have
\begin{align}
    D A_{\mu} = - \mathcal{L}_{\overset{\rm (D)}{\xi}}A_{\mu} = -\left(  \overset{\rm (D)}{\xi^{\rho}} \nabla_{\rho}A_{\mu}  + \nabla_{\mu}\overset{\rm (D)}{\xi^{\rho}}~A_{\rho} \right).
    \end{align}
 One can readily verify that the dS algebra commutation relations are satisfied, as \cite{Dobrev:1977qv} 
\begin{subequations}
\label{full algebra}
    \begin{align}
        [D,C_i]&={-C_i},~[D,T_i]=T_i~~,~~[T_i,C_j]={2D\delta_{ij}-2M_{ij}}~~\\
        [M_{ij},T_k]&=\delta_{ik}T_j-\delta_{jk}T_i,~~\\
        [M_{ij},C_k]&={\delta_{ik}C_j-\delta_{jk}C_i}~\\
        [M_{ij},M_{kl}]&= (-\delta_{jk} M_{il} - \delta_{il} M_{jk}) - (i \leftrightarrow j).
    \end{align}
\end{subequations}
%
These generators must be anti-hermitian when the representation under consideration is unitary.

To confirm the values of the scaling dimensions $\Delta_{\alpha}$ and $\Delta_{\beta}$, let us start our discussion from the bulk. The bulk field $A_{j}$ transforms under an infinitesimal dilatation as  $\delta_{D} A_{\mu} = - \mathcal{L}_{\overset{\rm (D)}{\xi}}A_{\mu}$, where $\overset{\rm (D)}{\xi} = \eta \partial_{\eta} + x^{i} \partial_{i}$.  Then, expanding the field in modes, we have 
    \begin{align}\label{eq:Lie deriv of quant field}
    \delta_{D}A_{j}(\vec{x},\eta) = -\sum_{\lambda=\pm}\int \frac{d^{3}k}{(2   \pi)^{3/2}} \left( a_{\vec{k},\lambda} ~ \mathcal{L}_{\overset{\rm (D)}{\xi}}A^{(\lambda,\vec{k})}_{j}(\vec{x},\eta)+ a^{\dagger}_{\vec{k},\lambda} ~ \mathcal{L}_{\overset{\rm (D)}{\xi}}A^{(\lambda,\vec{k})}_{j}(\vec{x},\eta)^{*} \right).
\end{align}
Also, by using the explicit expressions for the Bunch-Davies modes, we find
\begin{align}\label{dilatation on Bunch-Davies mode}
    \mathcal{L}_{\overset{\rm (D)}{\xi}}A^{(\lambda,\vec{k})}_{j}(\vec{x},\eta) = \left(i~k~|\eta|+i \vec{k}\cdot \vec{x}+1   \right) A^{(\lambda,\vec{k})}_{j}(\vec{x},\eta),
\end{align}
where we used $\delta_{D} A_{\mu} = - (\overset{\rm (D)}{\xi}+1)A_{\mu}$ (this follows from the standard formula for the Lie derivative). It is also easy to check that $ \mathcal{L}_{\overset{\rm (D)}{\xi}}A^{(\lambda,\vec{k})}_{\eta}(\vec{x},\eta)=0$ and $\partial^{j} \mathcal{L}_{\overset{\rm (D)}{\xi}}A^{(\lambda,\vec{k})}_{j}(\vec{x},\eta)=0$, i.e. dilatations preserve our gauge conditions.
Motivated by the form of the field at late times (\ref{eqn:Alower latetime_v2}), we define the infinitesimal transformation of the late-time operators $\alpha_j(\vec{k}), \beta_j(\vec{k})$ under dilatations through the following equation:
  \begin{align}
       \lim_{|\eta|\to0} \delta_{D}A_j(\vec{x},\eta)&\equiv\int\frac{d^3k}{(2\pi)^{3}}\Bigg[|\eta|^0~\delta_{D}\alpha_j(\vec{k})+|\eta| ~\delta_{D}\beta_j(\vec{k})\Bigg]e^{i\vec{k}\cdot\vec{x}}. 
\end{align}
To infer the infinitesimal transformations $\delta_{D}\alpha_j(\vec{k})$ and $\delta_{D}\beta_j(\vec{k})$, we take the late-time limit of (\ref{eq:Lie deriv of quant field}) explicitly, and we find 
    \begin{align}
       \lim_{|\eta|\to0} \delta_D A_j(\vec{x},\eta)&=\int\frac{d^3k}{(2\pi)^{3}}\Bigg[|\eta|^0 ~\alpha_j(\vec{k})(-x^{i}\partial_{i}-1)+|\eta|\beta_j(\vec{k})(-x^{i}\partial_{i}-2)\Bigg]e^{i\vec{k}\cdot\vec{x}}. 
\end{align}
It is easy to re-express this as follows
  \begin{align}
       \lim_{|\eta|\to0} \delta_D A_j(\vec{x},\eta)&=\int\frac{d^3k}{(2\pi)^{3}}e^{i\vec{k}\cdot\vec{x}}\Bigg[|\eta|^0\left(~(k^{i}\frac{\partial}{\partial k^{i}}+2)\alpha_j(\vec{k})\right)+|\eta|\left((k^{i}\frac{\partial}{\partial k^{i}}+1)\beta_j(\vec{k})\right)\Bigg],
\end{align}
where we integrated by parts in momentum space.
We thus find that the momentum-space late-time operators transform under infinitesimal  dilatations as
\begin{align}\label{dilatation transform of late-time operators}
    &\delta_{D} \alpha_j(\vec{k}) =(k^{i}\frac{\partial}{\partial k^{i}}+2)\alpha_j(\vec{k}), \nonumber \\
    & \delta_{D}\beta_j(\vec{k})=(k^{i}\frac{\partial}{\partial k^{i}}+1)\beta_j(\vec{k}).
\end{align}
Since these are dilatations in momentum space, the scaling dimensions should be interpreted as follows 
\begin{align}
    &\delta_{D} \alpha_j(\vec{k}) =\left(k^{i}\frac{\partial}{\partial k^{i}}+3-\Delta_{\alpha} \right)\alpha_j(\vec{k}),\\
    & \delta_{D}\beta_j(\vec{k})=\left(k^{i}\frac{\partial}{\partial k^{i}}+3-\Delta_{\beta} \right)\beta_j(\vec{k}),
\end{align}
in agreement with $\Delta_{\alpha}=1$ and $\Delta_{\beta}=2$. 
\section{Bulk QFT vs. late-time inner product}
\label{sec:Warm up: The QFT Hilbert space}

\subsection{The bulk QFT inner product and Hilbert space}

In section \ref{sec:Abelian gague field},  we introduced our canonically quantized vector field in the bulk. We decomposed it in  modes, introducing annihilation and creation operators. These operators satisfy \eqref{eqn:a ad intro}, which we write here  again for convenience, \
\begin{subequations}\label{eqn:a ad intro2}
\begin{align}
\label{annihilation and creation}\ak|0\rangle=0,&~~~~\adk|0\rangle=|\lambda,\vec{k}\rangle,\\
\label{eqn2:a adagger commutation}\left[a_{\vec{k},\lambda},a^\dagger_{\vec{k}',\lambda'}\right]&=(2\pi)^3~\delta_{\lambda\lambda'}\delta^{(3)}\left(\vec{k}-\vec{k}'\right).
\end{align}
\end{subequations}
One obtains single-particle states, $|\lambda,\vec{k}\rangle$, labeled by specific momentum $\vec{k}$ and helicity $\lambda \in \{ +,-\}$, from the action of the creation operator  on the vacuum, as in \eqref{annihilation and creation}. Single-particle states are (delta-function) normalized with respect to the QFT inner product,
\begin{align}
    \label{eq:qft inner product} \langle\lambda',\vec{k}'|\lambda,\vec{k}\rangle=~(2\pi)^3\delta^{(3)}\left(\vec{k}'-\vec{k}\right)~\delta_{\lambda' \lambda}.
\end{align}
 Note that hermitian conjugation $\dagger$ is defined with respect to this inner product, and thus,
\begin{align}
\nonumber  \langle\lambda,\vec{k}|= \left( |\lambda,\vec{k}\rangle\right)^\dagger=\langle 0| a_{\vec{k},\lambda}.
\end{align}

Single-particle states furnish the single-particle Hilbert space on which dS symmetries act. 
For each Killing vector in eq. (\ref{Killing vectors}) there exists a (hermitian) quantum dS charge $Q(\xi)$ which generates the corresponding infinitesimal dS transformation by acting on states, such as $\adk|0\rangle$, in the Hilbert space -- see, e.g. \cite{Cotaescu:2008hv}. As is well-known, conserved (Noether) charges $Q(\xi)$ can be expressed in terms of creation and annihilation operators
\begin{align}\label{hermitian quantum dS charges}
Q(\xi)=\left(Q(\xi)\right)^{\dagger} = -i~\sum_{\lambda = \pm}~\int \frac{d^3p}{(2 \pi)^3}a^{\dagger}_{\vec{p},\lambda}~\delta_\xi a_{\vec{p},\lambda},
\end{align}
where $\delta_\xi $ denotes the infinitesimal anti-hermitian transformation corresponding to the Killing vector $\xi$. It is easy to check that $Q(\xi)$ generates dS transformations, as
\begin{align}
    [a_{\vec{k},\lambda}, Q(\xi)] =-i~\delta_{\xi} a_{\vec{k},\lambda},~~~~~[a_{\vec{k},\lambda}^{\dagger}, Q(\xi)] =-i~\delta_{\xi} a_{\vec{k},\lambda}^{\dagger}
\end{align}
 Here we refer only to infinitesimal dS transformations in the bulk. The reason is that the coordinate system (\ref{conformal planar}) covers only half of the dS manifold, and finite special conformal transformations do not preserve the planar patch (\ref{conformal planar}).

Let us briefly explain how explicit expressions for the transformations $\delta_{\xi} a_{\vec{k},\lambda}$ can be found.
The transformation $\delta_\xi$ of the quantum field is attributed either to the mode solutions or to the operators $a_{\vec{k}, \lambda}$ and $a_{\vec{k}, \lambda}^{\dagger}$, as
\begin{align}
    \delta_{\xi}A_{j}(\vec{x},\eta) &= \sum_{\lambda=\pm}\int \frac{d^{3}k}{(2   \pi)^{3/2}} \left(  \delta_{\xi}a_{\vec{k},\lambda} ~ A^{(\lambda,\vec{k})}_{j}(\vec{x},\eta)+  \delta_{\xi}a^{\dagger}_{\vec{k},\lambda} ~ A^{(\lambda,\vec{k})}_{j}(\vec{x},\eta)^{*} \right)\\
    & = \sum_{\lambda=\pm}\int \frac{d^{3}k}{(2   \pi)^{3/2}} \left(  a_{\vec{k},\lambda} ~(- \mathcal{L}_{\xi}A^{(\lambda,\vec{k})}_{j}(\vec{x},\eta))+  a^{\dagger}_{\vec{k},\lambda} ~ (-\mathcal{L}_{\xi}A^{(\lambda,\vec{k})}_{j}(\vec{x},\eta)^{*}) \right).
\end{align}
One can show that the expression in the first line is equal to the expression in the second line by determining the transformation $\delta_{\xi}a_{\vec{k},\lambda} $. This can be verified by noting that the transformation $\delta_\xi$ of  annihilation (or creation) operators is determined from the transformation of the mode solutions under the Lie derivative. In particular,  using (\ref{annihilation op using KG}), we find
\begin{align}
    \delta_{\xi}a_{\vec{k}, \lambda} =&-(2 \pi)^{3/2}~\left(    A^{(\lambda, \vec{k})}, \mathcal{L}_{{\xi}}A\right)_{KG} =(2 \pi)^{3/2}~\left(   \mathcal{L}_{\xi} A^{(\lambda, \vec{k})}, A\right)_{KG} , \label{trasnfrmt of annihilation op under any Kil vec}
\end{align}
for any Killing vector $\xi$.
Here we used that $\mathcal{L}_{\xi}$ is anti-hermitian (up to a boundary term) with respect to the inner product (\ref{eqn:KGinner prod})%
\footnote{This can be readily proved by using the conserved Klein-Gordon current -- see \cite{Higuchi:STSHS} for the proof in global dS spacetime.}.~%
 Note that the transformed mode  $\mathcal{L}_{\xi} A^{(\lambda, \vec{k})}_{\mu}$, for a Killing vector $\xi$, can be expressed in terms of other positive-frequency modes, and thus $\delta_{\xi}a_{\vec{k}, \lambda}$ is expressed in terms of other annihilation operators.  However, the case of special conformal transformations (\ref{Killing vectors}) is an exception. In this case, the transformed mode $\mathcal{L}_{\xi} A^{(\lambda, \vec{k})}_{\mu}$ is expressed in terms of positive frequency modes plus a divergence-free pure-gauge mode\footnote{This becomes clear by observing that
$\mathcal{L}_{\overset{{\rm (i-SCT)}}{\xi}} A^{(\lambda, \vec{k})}_{\eta} \neq 0$, which can be re-expressed as the time derivative of a scalar function.}
\begin{align}\label{pure gauge mode}
  A^{(\text{PG})}_{\mu} (\vec{x}, \eta)= \partial_{\mu}f(\vec{x}, \eta) ,~~~ \nabla^\mu A^{(\text{PG})}_{\mu} (\vec{x}, \eta)=0.
\end{align}
%
In fact, special conformal transformations preserve the Lorenz gauge $\nabla^{\mu}A_{\mu}=0$, while all other dS transformations preserve the Coulomb gauge $A_{\eta}=0=\nabla^{j}A_j$. However, divergence-free pure-gauge modes $ A^{(\text{PG})}_{\mu} (\vec{x}, \eta)$ are orthogonal to themselves and to all  physical modes (up to boundary terms)%
\footnote{This can be proved as follows. Consider the Klein-Gordon current $J^{\mu}_{KG}(A^{(1)}, A^{(2)})$ (\ref{eqn:KG current}). Let  $A^{(1)}_\nu$ be any solution in the Lorenz gauge and $A^{(2)}_\nu$ be a pure-gauge mode $A^{(2)}_\nu= A^{(PG)}_\nu = \partial_\nu f$ (\ref{pure gauge mode}) in the Lorenz gauge. A straightforward calculation shows that $J^{\mu}_{KG}(A^{(1)}, A^{(PG)}) = \nabla_{\nu}B^{\nu \mu}$ where $B_{\mu \nu} = -B_{\nu \mu}$ is an anti-symmetric tensor. Thus, from (\ref{KG product in terms of KG current}) we find $ (A^{(1)}, A^{(PG)})_{KG}= \int d^{3}x ~\partial_{j}(\sqrt{-g}B^{j \eta})$, which is a boundary term.}.~%
As a consequence, the appearance of pure-gauge modes in the case of special conformal Killing vectors in (\ref{trasnfrmt of annihilation op under any Kil vec}) does not affect the transformation of  creation/annihilation operators. 

To conclude, $\delta_{\xi}a_{\vec{k},\lambda} $ is expressed in terms of other annihilation operators for all Killing vectors $\xi$. This means that all quantum dS charges (\ref{hermitian quantum dS charges}) preserve the Bunch-Davies vacuum 
\begin{align}\label{dS invariance of Bunch-Davies vacuum}
    Q(\xi)|0 \rangle =0.
\end{align}
To facilitate the discussions given in the previous paragraph, some explicit expressions for the dS transformations of annihilation operators are in order.
\begin{itemize}
    \item  For an infinitesimal dilatation (\ref{Killing vectors}), using eq. (\ref{trasnfrmt of annihilation op under any Kil vec}) we find that annihilation operators transform as
\begin{align}
    \delta_{D}a_{\vec{k}, \lambda} 
    =&  \Big(k^{m}\frac{\partial}{\partial k^{m}}  + \frac{3}{2} \Big)a_{\vec{k}, \lambda} \nonumber \\
    =& \int d^{3}q~~~ \Bigg(\Big(-q^{m}\frac{\partial}{\partial q^{m}}  - \frac{3}{2} \Big)\delta^{(3)}(\vec{q} - \vec{k})\Bigg)~a_{\vec{q}, \lambda},
    \label{trasnfrmt of annihilation op dilatation}
\end{align}
where we also used the transformation formula for the mode solution (\ref{dilatation on Bunch-Davies mode}). Integrating by parts, one can show that the second line in (\ref{trasnfrmt of annihilation op dilatation}) is equal (up to boundary terms) to the first. As a consistency check, note that using eq. (\ref{trasnfrmt of annihilation op dilatation}) we can re-derive the dilatation transformation laws of late-time operators (\ref{dilatation transform of late-time operators}).

\item For an infinitesimal translation (\ref{Killing vectors}) in the $j$-direction, we find
\begin{align}
    \mathcal{L}_{\overset{{\rm (j-Trans)}}{\xi}}A^{(\lambda,\vec{k})}_{\mu} = i\,k_j\,A^{(\lambda,\vec{k})}_{\mu},
\end{align}
and thus,
\begin{align} \label{trasnfrmt of annihilation op translation}
    \delta_{\overset{{\rm (j-Trans)}}{\xi}}a_{\vec{k},\lambda}& = -i~k_j~a_{\vec{k},\lambda} \nonumber\\
    &= -i~\int d^3q ~\delta^{(3)}(\vec{q}-\vec{k})~q_{j}~a_{\vec{q},\lambda}.
\end{align}
Expressions similar to (\ref{trasnfrmt of annihilation op dilatation}) and (\ref{trasnfrmt of annihilation op translation}) can be also found for special conformal transformations and rotations.

\item To sum up, an infinitesimal transformation of annihilation/creation operators generated by a generic Killing vector $\xi$ (\ref{Killing vectors}) is expressed as 
\begin{align} 
&\delta_{\xi}a_{\vec{k},\lambda}
    = ~\int d^3q~\sum_{\lambda'=\pm} ~m_{\xi}(\vec{k},\lambda | \vec{q},\lambda' )~a_{\vec{q},\lambda'}, \nonumber\\
    &\delta_{\xi}a^{\dagger}_{\vec{k},\lambda}
    = ~\int d^3q ~\sum_{\lambda'=\pm}~m_{\xi}(\vec{k},\lambda | \vec{q},\lambda' )^{*}~a_{\vec{q},\lambda'}^{\dagger},~~\lambda,\lambda' \in \{ +,-\},
\end{align}
where the coefficients $m_{\xi}(\vec{k},\lambda | \vec{q},\lambda')$ are the ``matrix elements'' of the corresponding generator.
The anti-hermiticity of $\delta_\xi$ means that
\begin{align}
    \langle0|{a}_{\vec{k}', \lambda'}~(\delta_{\xi}{a}^{\dagger}_{\vec{k}, \lambda}) |0\rangle +  \langle0|(\delta_{\xi}{a}_{\vec{k}', \lambda'})~{a}^{\dagger}_{\vec{k}, \lambda} |0\rangle =0,
\end{align}
which implies that
\begin{align}
    m_{\xi}\left(\vec{k},\lambda|  \vec{k}', \lambda' \right)^{*} = -  m_{\xi}\left(\vec{k}', \lambda'|\, \vec{k} , \lambda\right).
\end{align}
 In practice, the coefficients $m_{\xi}\left(\vec{k},\lambda|  \vec{k}', \lambda' \right)$ vanish for $\lambda ' \neq \lambda$, and we may write $m_{\xi}\left(\vec{k},\lambda|  \vec{k}', \lambda' \right) \equiv ~\delta_{\lambda  \lambda'}~m^{(\lambda)}_{\xi}\left(\vec{k}|  \vec{k}'\right)$. That is, creation (or annihilation) operators with fixed helicity label $\lambda$  transform only among themselves under $so(4,1)$ -- helicity does not change under dS transformations. This will become clear from the discussions in sections \ref{sec:field strength and anti-self-duality} and \ref{sec:Direct sum of representations in the Hilbert space}.
\end{itemize}

\subsection{Late-time  inner product (warm-up)}

In the following sections, we will consider how states  arising from the action of the late-time operators $\alpha_i$ and $\beta_i$ (identified in section \ref{sec:Abelian gague field}) are normalizable with a specific choice of a CFT-inspired inner product. Note that although late-time operators are also built out of $\ak$ and $\adk$, they are further equipped with polarization tensors and extra factors  of $k^{\pm \frac{1}{2}}$. In particular, we will identify a CFT-inspired \cite{Dobrev:1977qv} ``late-time inner product'', which is $SO(4,1)$-invariant, and appropriate for the discrete series. In particular, the following ``late-time states'',
\begin{subequations}\label{eqn:intro alpha beta kets}\begin{align}
|\alpha_i(\vec{k})\rangle&\equiv\alpha_i(\vec{k})|0\rangle,\\
|\beta_i(\vec{k})\rangle&\equiv\beta_i(\vec{k})|0\rangle,
\end{align}\end{subequations}
will be shown to be normalizable with respect to this inner product. Using such a CFT-inspired inner product for late-time operators is natural since these operators behave as conformal primaries with respect to the conformal group of the late-time boundary, which is none other than the dS group $SO(4,1)$.
 Similar late-time inner products for scalar fields (in any number of dimensions) have been discussed in \cite{unitarity,twopoint, Sengor:2023buj}.

\section{Exceptional series inner product and late-time operators}
\label{sssec:The corresponding discrete series late-time operators}
In section \ref{sec:Brief summary of discrete series in four spacetime dimensions}, we start with a short summary of the general features of exceptional series representations of $SO(d+1,1)$, where we mainly follow \cite{Dobrev:1977qv}. We also identify the exceptional series function spaces%
\footnote{We use the term `function space' in order to follow the terminology of \cite{Dobrev:1977qv}. For the purposes of our work, this term can be used interchangeably with `representation space'. } 
that host our late-time operators. We discuss the normalization of our operators with respect to the appropriate exceptional series inner products in section \ref{subsec:The discrete series inner product for beta}. 
Our main result in this section are as follows. The late-time operator $\alpha_j$ creates states with  finite positive norm with respect to the CFT-inspired exceptional series inner product (see section \ref{sec:Warm up: The QFT Hilbert space}), furnishing the unitary discrete series representation of $SO(4,1)$ associated with the photon on $dS_4$. In the case of $\beta_j$, by exploiting a known isomorphism of $SO(4,1)$ representation spaces -- and ``importing''  the inner product from $\alpha_j$ -- we find a realization of the same  unitary photon  representation as in the case of 
$\alpha_j$.

\subsection{Brief summary of exceptional series}
\label{sec:Brief summary of discrete series in four spacetime dimensions}

For all dimensions $d$, there exist principal and complementary series of unitary and irreducible representations of $SO(d+1,1)$.  There is also another series of representations called \emph{exceptional series}. When $d+1$ is even there also exist discrete series representations. For $d+1=4$, which is our case of interest, exceptional series coincide with discrete series. Below we give some details on the exceptional series.

The exceptional series of $SO(d+1,1)$ host four subcategories of representations that share the same value for their quadratic Casimir and are, in general, not irreducible, while not all of them are unitarizable. See the first two columns of Table \ref{exceptional table1} for the list of the four different categories of exceptional series representations of $SO(d+1,1)$. See also \cite{sun2021note, Dobrev:1977qv} for details on unitarity and on the quotient space method needed to construct irreducible exceptional series representations -- the irreducible representations are called `invariant subspaces' in \cite{Dobrev:1977qv}, and they are given in the last column in Table \ref{exceptional table1}.

\begin{table}[h]
\begin{center}
\begin{tabular}{|c|c|c|}
\hline
Exceptional series category & function space & invariant subspace\\
\hline
$\chi^-_{l_\sigma}=\{l,1-\frac{d}{2}-l-\sigma\}$ & $C^-_{l\sigma}$ & $E_{l\sigma}$\\
\hline
$\chi^+_{l_\sigma}=\{l,\frac{d}{2}+l+\sigma-1\}$ & $C^+_{l\sigma}$ & $F_{l\sigma}$\\
\hline
$\chi^{'-}_{l_\sigma}=\{l+\sigma,1-\frac{d}{2}-l\}$ & $C^{'-}_{l\sigma}$ & $F'_{l\sigma}$\\
\hline
$\chi^{'+}_{l_\sigma}=\{l+\sigma,\frac{d}{2}+l-1\}$ & $C^{'+}_{l\sigma}$ & $D_{l\sigma}$\\
\hline	
	\end{tabular}
 \end{center}
\caption{\label{exceptional table1}The four cases of exceptional series representation spaces (second column). The corresponding representation-theoretic labels are given in the first column. The last column includes the irreducible exceptional series representations (only three of them are inequivalent, while not all of them are unitarizable) \cite{sun2021note, Dobrev:1977qv}.}
 \end{table}

 The different types of the exceptional series representations can be related to each other by specific transformations  (similarity transformations), realized by the \emph{intertwining operators}%
 \footnote{In the language of \cite{Dobrev:1977qv}, an intertwining operator $A$, maps two continuous representations to each other, as in equation (3.1) of \cite{Dobrev:1977qv}
 \begin{align}
     \label{eq:intertwining op}A T(g)=T'(g)A,
 \end{align}
 for all group elements $g$, where $T(g)$ and $T'(g)$ are two continuous representations. When the intertwining operator $A$ has a continuous inverse, relation \eqref{eq:intertwining op} becomes a similarity transformation. It is such invertible intertwining operators that we are after.}
 that change the \emph{scaling dimension} -- details on the scaling dimension will be given below. The intertwining operators are normalizable and invertible operators that are well-defined under very specific conditions \cite{Dobrev:1977qv}. The well-defined intertwining operators  play an important role in the definition of the inner product that renders  the representation unitary. {Below, we discuss the representation-theoretic labels relevant to $SO(d+1,1)$ and demonstrate that our late-time operators correspond to the exceptional series.}  

 \paragraph{Representation-theoretic labels (general details).} The eigenvalue of the quadratic Casimir  informs us about the labels of the representations.  In the case of dS group $SO(d+1,1)$ (or its algebra $so(d+1,1)$), \emph{spin} and \emph{scaling weight} are the representation-theoretic labels that appear in the eigenvalue of the quadratic Casimir%
 \footnote{To be precise, the eigenvalue of the quadratic Casimir depends on the scaling weight and on the $SO(d)$ highest weight. For example for a totally symmetric tensor with $s$ indices, the  $SO(d)$ highest weight is $(s,0,...,0)$. For $d=3$, the $SO(3)$ highest weight is simply given by one number: the spin $s$. For more details, see, e.g. \cite{Dobrev:1977qv, Basile:2016aen}.}.
We will denote \emph{spin}  by $s$ and we will shortly see that, for some of the representations, $s$ itself coincides with the label $l$ of the compact rotation subalgebra $so(d)$%
\footnote{In two spacetime dimensions spin is replaced by parity of evenness and oddness of functions that realize the unitary irreducible representation \cite{gelfand1966generalized}.}.~%
The scaling weight, $c$, can be expressed in terms of the scaling dimension $\Delta$. In particular, as coordinates are scaled by a parameter $\kappa \in \mathbb{R}$, operators of scaling dimension $\Delta$ transform as follows
\begin{align}
    \label{scaling dim intro} \mathcal{O}\left(\kappa\vec{x}\right)=\kappa^{-\Delta}\mathcal{O}\left(\vec{x}\right)~~\text{as}~~ x^i\to\kappa x^i,
\end{align}
where  $\vec{x} \in \mathbb{R}^{d}$, with
\begin{align}
    \label{c intro} \Delta=\frac{d}{2}+c.
\end{align}
The term $\frac{d}{2}$ here, plays the role of half sum of the restricted positive roots \cite{Dobrev:1977qv}.
For a totally symmetric, transverse-traceless integer-spin field on $dS_{d+1}$, the labels $s,\Delta$ and $c$ appear in the $SO(d+1,1)$ quadratic Casimir eigenvalue as
\begin{align}
    Casimir&= \Delta(\Delta-d)+s(s+d-2)\\
           &=c^2-\frac{d^2}{4}+s(s+d-2).
\end{align}

 From a field-theoretic perspective, the scaling weight includes information about the mass of the field\footnote{{Here by mass we mean the mass parameter that appears in the free field equation. The concept of physical mass in dS is subtle, see e.g. \cite{Gazeau:2024vfe} for a relevant  discussion.}}. Depending on how light or heavy the mass is, $c$ can be real or purely imaginary. Both real and imaginary values are allowed by unitarity and different ranges on the scaling weight correspond to different categories of representations. The fact that complex scaling dimensions can correspond to unitary irreducible representations is a feature of the dS group $SO(d+1,1)$ that sets it apart from the Poincaré ($ISO(d,1)$) and AdS ($SO(d,2)$) groups. This feature is linked with the lack of global time translations as isometries in dS, while they exist as isometries for the other two spacetimes.  Although not an exhaustive list of references on the subject, we refer the reader to \cite{Dobrev:1977qv,Basile:2016aen,sun2021note,Hinterbichler:2026xqf} for comprehensive reviews, to \cite{unitarity,twopoint,LT14proceed,Corfu2022proceed,particles,Anninos:2023lin, Higuchi:1991tn, Higuchi:STSHS} for examples on realizations of different categories of representations in scalar field theoretic examples and to \cite{Letsios:2023qzq, Letsios:2022tsq, Letsios:2023awz,Letsios:2025pqo, Higuchi-Letsios, Anninos:2025mje} for field theoretic realizations regarding fermions. 

\paragraph{Exceptional series: irreducibility.} To sum up, the representations of $SO(d+1,1)$  are collectively labeled by 
\begin{align}\label{dSlabels intro} \chi=\{s,c\},\end{align}
while for specific values of $s$ and $c$ the representations are unitary.
The exceptional series representations involve two additional labels, $l$ and $\sigma$, which take the following values (for the cases of integer-spin representations) \cite{Dobrev:1977qv}
\begin{subequations}\label{l nu ranges}\begin{align}
    l&=0,1,2,\dots\\
    \sigma&=1,2,3,\dots.
\end{align} 
\end{subequations}
The physical meaning of these labels, in the context of dS field theory, will be discussed below.  As mentioned earlier, the four exceptional series representations in the second column of Table \ref{exceptional table1} are, in general, not irreducible. There are three inequivalent irreducible exceptional series representations of $SO(d+1,1)$ (i.e. invariant subspaces in Table \ref{exceptional table1}): $F_{l\sigma} \cong F_{l\sigma}'$,   $D_{l\sigma}$ and $E_{l\sigma}$ \cite{Dobrev:1977qv}. Note the following isomorphisms \cite{Dobrev:1977qv} 
\begin{align} \label{isomorphisms}
    F_{l\sigma} \cong F_{l\sigma}' \cong C^{-}_{l \sigma} /E_{l \sigma} \cong C^{'+}_{l \sigma}/D_{l \sigma},~~~~E_{l \sigma}=C^{+}_{l \sigma}/F_{l \sigma},~~~~D_{l \sigma}\cong C^{'-}_{l \sigma}/F^{'}_{l \sigma}.
\end{align}
\paragraph{Note.} The case $d+1=4$ is special as exceptional series coincide with the discrete series, and in this case  $D_{l_\sigma}$ can be reduced further. In particular, for $d+1=4$, $D_{l \sigma}$ is isomorphic to the direct sum of two irreducible discrete series representations that are the mirror image of each other:
\begin{align}\label{discrete series in 4D split into two}
    \text{for}~~d=3,~~~~~~~~D_{l \sigma} \cong D_{l \sigma}^{+} \oplus D_{l \sigma}^{-}.
\end{align}

\paragraph{Exceptional series: irreducibility and unitarity.}  Not all irreducible representations in (\ref{isomorphisms}) are unitary, i.e. not all of them admit a positive-definite and $SO(d+1,1)$-invariant inner product. The unitary irreducible exceptional series representations of $SO(d+1,1)$ are the \cite{Dobrev:1977qv, ottoson1968classification, schwarz1971unitary}%
\footnote{For convenience, we do not discuss fermionic representations here. See \cite{Hinterbichler:2026xqf, Letsios:2022tsq, Letsios:2023qzq, Letsios:2023awz, Letsios:2025pqo, Anninos:2025mje} for discussions on fermionic representations of $SO(d+1,1)$.} %
: \emph{exceptional series-I} ($F_{0 \sigma}$)  and \emph{exceptional series-II} ($D_{l \sigma}$).

The unitary exceptional series-I capture only scalar fields on $dS_{d+1}$ with specific values for their mass, $\Box \phi = m^{2}\phi$,
\begin{equation}
    \text{exceptional series-I,}~F_{0 \sigma}:~~s=0,~ m^2=-(\sigma-1)(d+\sigma-1)=\Delta(d-\Delta).
\end{equation}
These correspond to `tachyonic' shift-symmetric scalars among which the massless (with $\sigma=1$) scalar is also included \cite{Higuchi:STSHS, Bros:2010wa}. In previous work, we have investigated examples in the exceptional series-I - in particular, the massless scalars in two \cite{Anninos:2023lin, Sengor:2023buj} and four \cite{twopoint} dimensions.

Massless gauge fields, such as photons and gravitons, as well as partially massless ones, belong to exceptional series-II. The corresponding labels are:
\begin{equation} 
\text{exceptional series-II,}~D_{l \sigma}:~~s=l+\sigma,~~l=0,1,\dots, s-1,~~\sigma=1,2,\dots,s.
\end{equation}
Here, $l$ is the `depth' of the (partially) massless field spin-$s$ field, where the value $l=s-1$ corresponds to the strictly massless case. Thus, the massless spin-1 field has depth $l=0$. For each value of the spin $s$ and depth $l$, the value of the parameter $\sigma$ is fixed as $\sigma = s-l$. The scaling dimension for the spin-$s$ (partially) massless field is
$\Delta=d+l-1$ (or $d-\Delta$).
In the case of $SO(4,1)$, exceptional series-II coincide with the discrete series, and they correspond to a direct sum of two unitary irreducible representations describing fields with  opposite helicity. This connection to helicity will become clear in the following sections.

\paragraph{Identifying exceptional series late-time operators.} 
Having obtained our vector field late-time operators on $dS_4$ and identified their scaling dimensions, we can readily identify their scaling weights, as
\begin{subequations}
    \begin{align}
        \label{c alpha} \Delta_\alpha&=1 ~~\implies~~ c_\alpha=-\frac{1}{2},\\
        \label{c beta} \Delta_\beta&=2~~\implies~~ c_\beta=\frac{1}{2}.
    \end{align}
\end{subequations}
Denoting the representation-theoretic labels for the late-time operators $\alpha_j$ and $\beta_j$ as $\chi^{\alpha}=\{s,c_\alpha\}$ and $\chi^{\beta}=\{s, c_{\beta}\}$, respectively, we have
\begin{subequations}
    \begin{align}
        \chi^\alpha=\left\{1,-\frac{1}{2}\right\},\\
        \chi^\beta=\left\{1,\frac{1}{2}\right\}.
    \end{align}
\end{subequations}
Moreover, we have $s=1$ for the spin, while the depth $l$ and the label $\sigma$ are given by
    \begin{align}\label{ l nu for s1} l=0,~~\sigma=s-l=1.
    \end{align}
Thus,
 \begin{align}
     \chi^\beta&=\left\{1,\frac{1}{2}\right\}=\chi^{'+}_{0_1},\\
     \chi^\alpha&=\left\{1,-\frac{1}{2}\right\}=\chi^{'-}_{0_1},
 \end{align}
 see Table \ref{exceptional table1}.

\subsection{Normalizing states at the late-time}
\label{subsec:The discrete series inner product for beta}

Following \cite{Dobrev:1977qv}, in this section we identify the appropriate group-theoretic (CFT-inspired) inner products in the space of states created by our late-time operators. We find that the late-time operator $\alpha_i$, which is the leading one at late-times, gives rise to states that are normalizable with respect to the appropriate exceptional series inner product. These states furnish the unitary discrete series representation of $SO(4,1)$ associated with the photon (that is $D_{01}$ in Table \ref{exceptional table1} and eq. (\ref{discrete series in 4D split into two})). We also discuss how the same unitary representation can be realized in terms of states created by $\beta_j$.

\paragraph{Background material.}
 Our late-time operators give a realization of representations of $SO(4,1)$. The inner product used for the normalization of the states created by the late-time  operators must remain invariant under the action of the group. A standard method to construct invariant inner products concerns the realization of $SO(4,1)$ representations on spaces of functions of group elements \cite{Dobrev:1977qv}. These are functions of $SO(4,1)$ group elements and they form the Hilbert space which realizes the unitary representations of the rotation subgroup $SO(3)$. Then, in the case of unitary representations, the action of the group transformations on the aforementioned functions over group elements should preserve the inner product on these functions \cite{Dobrev:1977qv}.  The function spaces of interest come with their set of defining properties, which encode the action of group elements.  There is a well-defined method to relate functions of group elements to functions in position space, as described in section 2D of \cite{Dobrev:1977qv},  where translations play the key role%
 \footnote{ These are translations in space, which form the  subgroup $\tilde{N}$  of $SO(d+1,1)$. There is a specific way to relate a point  $\vec{x}\in\mathbb{R}^d$ to a particular translation element $\tilde{n}_{\vec{x}}\in \tilde{N}$ which can be used to map functions of translation elements to functions in position space (this is given by equation (2.26) of \cite{Dobrev:1977qv}, as well as equation (3.15) of \cite{unitarity}). Once a translation element has been associated with a particular point in position space, there is a further relation that relates a generic group element $\in SO(d+1,1)$ to a particular point in position space  (see equation (2.27) of \cite{Dobrev:1977qv}, as well as equation (3.17) of \cite{unitarity}). Using the aforementioned relations, we can re-express the inner product for functions over group elements to an inner product for functions in position space. Note that the integration measure transforms non-trivially (for details see \cite{unitarity} section 4).  }%
 - see also \cite{unitarity}. This relation allows one to study the same representation realized on functions in position space, and thus, work with an inner product for  functions in position space. One can also work in momentum space by performing a Fourier transformation. In the present paper we work with inner products of states in momentum space.

As we explicitly worked out in \cite{unitarity} (section 4.1), following section 3D of \cite{Dobrev:1977qv}, when one considers the action of $SO(d+1,1)$ on the inner product (in the realization where the representation space consists of functions of group elements), the dilatation part of the group transformation gives rise to a  factor of $|a|^{(c+c^*)}$, where $a$ is a dilatation element, $^*$ stands for complex conjugation, and $c$ is the scaling weight. To ensure the invariance of the inner product under the action of the group, this  factor must be ``eliminated''. To achieve this, one introduces  an invertible, normalizable operator that generates a similarity transformation of the representation. This similarity transformation is such that it sends $c\to-c$ while keeping the spin label the same. Because the effect of this similarity transformation is to send the scaling dimension $\Delta$ to $d-\Delta$, this is referred to as a shadow transformation in  the CFT literature. The operator that encodes this shadow transformation, which can be thought of as a very specific similarity transformation, is called an \emph{intertwiner} {(or \emph{intertwining operator})}. We will denote the intertwiner by $G$ following the notation of \cite{Dobrev:1977qv}, and we will also introduce some extra labels depending on the representation of interest. 

The inner product itself involves complex conjugation of states. In the case of principal series, the factor of $|a|^{(c+c^*)}$ automatically disappears, because the scaling weight $c$ is purely imaginary. The intertwiner in this case is just the identity. The presence of the intertwiner is more crucial for the rest of the categories of representations, as they have real $c$. The normalizability and invertibility of the intertwiner work differently for different ranges of $c$: the intertwiner applicable to complementary series is not the same as that applicable to exceptional and discrete series. 

A detailed review of the intertwiner for the case of bosons, the function spaces on which it acts, as well  as its explicit form can be found in \cite{Dobrev:1977qv}, which is our main reference. More recent reviews  are \cite{sun2021note,Schaub:2024rnl}. Reference \cite{Schaub:2024rnl} also gives a very pedagogical summary of the method of harmonic analysis involved, as well as a discussion on fermionic representations. Here we will only refer to the intertwiner we need to make use of. Discussions concerning the case of scalars, along with a  detailed review, can be found in \cite{unitarity,twopoint,particles}.

\paragraph{Translating our notation to the notation of \cite{sun2021note}.} In the present paper, we mainly follow the notation of \cite{Dobrev:1977qv}.  For reasons of clarity, note that in the notation of \cite{sun2021note} the function space $C'^+_{0_1}$ corresponds to $\mathcal{F}_{d+t-1,s}|_{d=3,t=0,s=1}=\mathcal{F}_{2,1}$. The intertwiner $G'^-_{0_1}$ that acts on this function space corresponds to $S^{-}_{2,1}$ in \cite{sun2021note}.  Also, the function space $C'^-_{0_1}$ corresponds to  $\mathcal{F}_{1-t,s}|_{t=0,s=1}=\mathcal{F}_{1,1}$ in the conventions of \cite{sun2021note}.  The intertwining operator that acts on this function space is $G'^+_{0_1}$, which corresponds to $S^+_{1-t,s}|_{t=0,s=1}=S^+_{1,1}$ in the notation of \cite{sun2021note}. More details are given below.

\paragraph{Identifying the intertwiner and inner product for $\alpha_j(\vec{k})$.} Under infinitesimal dilatations (in momentum space), the $\Delta=1$ late-time operator $\alpha_j$ transforms as shown in (\ref{dilatation transform of late-time operators}). Thus, the  representation-theoretic labels for  $\alpha_j(\vec{k})$  are given by $\chi'^-_{0_1}$ in Table \ref{exceptional table1}, with corresponding function space $C'^-_{0_1}$.   The intertwining operator  $G'^+_{0_1}$ acts on this space. From \cite{sun2021note}, equation (4.66),\footnote{The convention is such that $$(a)_k=\frac{\Gamma(a+k)}{\Gamma(a)}.$$} we have
\begin{align}\label{defining the intertwiner S_1,1^+}
   G'^+_{0_1}(\vec{k})\equiv S^+_{1,1}(\vec{k})=k~\sum^{s=1}_{l=1}\frac{(l)_{1-l}}{(1+l)_{1-l}}\Pi^{1\,l}(\hat{k})=k~\Pi^{11}(\hat{k}),
\end{align}
where $\Pi^{1\,1}$ is one of the projection operators $\Pi^{lm}$ ($S^+_{1,1}(\vec{k})$ is the notation used in \cite{sun2021note}). The projection operators $\Pi^{lm}$ project spin-$l$  representations of $SO(3)$ into spin-$m$  representations of $SO(2)$.  The  construction of intertwiners is translation-invariant and  $SO(2)$ can be conveniently viewed as the  stability group of a fixed reference momentum 3-vector\footnote{For example, one can pick the direction of propagation as the third direction and the reference momentum vector can be chosen as
\begin{align}\label{eqn:khat}
    \vec{k}=\begin{pmatrix}
        0\\
        0\\
        |k|
    \end{pmatrix}=|k|\hat{k}.
\end{align}}. In some sense, this is the dS analogue of the Little Group when discussing representations of the Poincaré group. For discussions on the defining properties of  projection operators see \cite{Dobrev:1977qv,Schaub:2024rnl}.%
\footnote{
In the various existing reviews \cite{Dobrev:1977qv, Schaub:2024rnl, sun2021note}, some of the intertwining operators are given in ambient space and in index-free formalism. In the index-free formalism, indices   are  contracted with null vectors and one works with polynomials of null vectors instead of tensors. One can `free' the indices by acting with the Todorov operator, which is also known as the `interior derivative'. The operator that frees the indices is a differential operator that differentiates polynomials with respect to null vectors in a subtle way, to give back tensors with free indices. The index free formalism is reviewed in \cite{Penedones:2023uqc}, as well as in \cite{Pethybridge:2021rwf} with more emphasis on its use for dS spacetime. In the present paper, we  use expressions with explicit indices. Ref. \cite{sun2021note} provides some of the intertwining operators with their explicit indices.  }
The projection operator $\Pi^{11}$ is given by \cite{sun2021note}
\begin{align}
    \left[\Pi^{11}(\hat{k})\right]_{ij}=\delta_{ij}-\hat{k}_i\hat{k_j}.
\end{align}
The action of the intertwining operator  $ G'^+_{0_1}$ on the late-time operator $\alpha_j$ gives the shadow operator $\tilde{\alpha}_i$ with scaling dimension $\bar{\Delta}_\alpha=3-\Delta_\alpha=2$, as
\begin{align}\label{eqn:tildealpha}
    \tilde{\alpha}_i(\vec{k})=\left[G^{'+}_{01}(\vec{k})\right]_{i}\,^j~\alpha_j(\vec{k}).
\end{align}
Recalling  that
\begin{align}
   \alpha_j(\vec{k}) =i\,{{\frac{1}{\sqrt{2}}}}\sum_{\lambda=\pm}\hat{\epsilon}_{j}^{(\lambda)}(\vec{k})\left[\admk-\ak\right]k^{-\frac{1}{2}},
\end{align}
we find an explicit expression for the shadow
\begin{align}
 \tilde{\alpha}_i(\vec{k})=i{{\frac{1}{\sqrt{2}}}}\sum_{\lambda=\pm}\hat{\epsilon}_{i}^{(\lambda)}(\vec{k})\left[\admk-\ak\right]k^{\frac{1}{2}}= ~k~{\alpha}_i(\vec{k}).
\end{align}
Note that the action of the intertwining operator changes the $k$-dependence.

Now we define the densitized inner product that can normalize the late-time operator $\alpha_j(\vec{k})$ as follows: 
\begin{align}\label{eqn:densitizedinner_alpha}\frac{1}{\Omega}\left(\alpha,~G^{'+}_{01}\alpha\right)& \equiv\frac{1}{\Omega}\int \frac{d^3k}{(2\pi)^3}\langle\alpha_i(\vec{k})|\tilde{\alpha}_i(\vec{k})\rangle ,
\end{align}
where
\begin{align}
\Omega\equiv\sum_{\lambda=\pm}\int\frac{d^3k}{(2\pi)^3}\langle-\vec{k},\lambda|-\vec{k},\lambda\rangle
\end{align}
is the volume of momentum eigenstates.  Here, $\langle\alpha_i(\vec{k})|\tilde{\alpha}_i(\vec{k})\rangle$  $\equiv \sum_{i \in \{x,y,z\}}\langle\alpha_i(\vec{k})|\tilde{\alpha}_i(\vec{k})\rangle$. Such an inner product is similar to the densitized inner product defined for the scalars given in \cite{unitarity}. The ``state operators'' $|\mathcal{O}(\vec{k})\rangle$ are defined by the action of the corresponding operator on the Bunch-Davies vacuum $|0\rangle$, as follows 
\begin{align}
|\tilde{\alpha}_i(\vec{k})\rangle&\equiv\tilde{\alpha}_i(\vec{k})|0\rangle\\
    &=i{{\frac{1}{\sqrt{2}}}}k^{\frac{1}{2}}\sum_{\lambda=\pm}\hat{\epsilon}_{i}^{(\lambda)}(\vec{k})|-\vec{k},\lambda\rangle,
    \end{align}
\begin{align}  |\alpha_i(\vec{k})\rangle&\equiv\alpha_i(\vec{k})|0\rangle\\
    &=i{{\frac{1}{\sqrt{2}}}}k^{-\frac{1}{2}}\sum_{\lambda=\pm}\hat{\epsilon}_{i}^{(\lambda)}(\vec{k})|-\vec{k},\lambda\rangle,
\end{align}
with  corresponding bra-state 
\begin{align}
  \langle\alpha_i(\vec{k})|=(|\alpha_i(\vec{k}) \rangle )^{\dagger}.
\end{align}
In order to determine the value of the densitized inner product  \eqref{eqn:densitizedinner_alpha}, we compute
\begin{align}
    \langle\alpha_i(\vec{k})|\tilde{\alpha}_i(\vec{k})\rangle&={\frac{1}{2}}\sum_{\lambda,\lambda'}\hat{\epsilon}_{i}^{(\lambda)}(\vec{k})\hat{\epsilon}_{i}^{(\lambda')}(\vec{k})^*\langle-\vec{k},\lambda'|-\vec{k},\lambda\rangle\\
    &={\frac{1}{2}}\sum_{\lambda}\langle-\vec{k},\lambda|-\vec{k},\lambda\rangle,
\end{align}
where we used eq. (\ref{eqn:pol normalization}) for the polarization vectors.
Thus,
\begin{align}
    \frac{1}{\Omega}\left(\alpha,G^{'+}_{01}\alpha\right)={\frac{1}{2}},
\end{align}
which is positive.

For convenience, let us also define the normalized late-time operator
\begin{align}\label{eqn:normalized alpha}
   \alpha^N_j(\vec{k}) =i\sum_{\lambda=\pm}\hat{\epsilon}_{j}^{(\lambda)}(\vec{k})\left[\admk-\ak\right]k^{-\frac{1}{2}},~~\Delta_\alpha=1.
\end{align}
The operator $\alpha^N_j(\vec{k})$ gives rise to states that furnish the unitary discrete series representation $D_{01} \cong C^{'-}_{0 1}/F^{'}_{0 1}$ (\ref{discrete series in 4D split into two}) of $SO(4,1)$, 
\begin{align}  |\alpha^N_i(\vec{k})\rangle&\equiv\alpha^N_i(\vec{k})|0\rangle\\
    &=ik^{-\frac{1}{2}}\sum_{\lambda=\pm}\hat{\epsilon}_{i}^{(\lambda)}(\vec{k})|-\vec{k},\lambda\rangle.
\end{align}
The normalized shadow operator
\begin{align}
\label{tildealphaN}
\tilde{\alpha}^N_i(\vec{k})=i\sum_{\lambda=\pm}\hat{\epsilon}_{i}^{(\lambda)}(\vec{k})\left[\admk-\ak\right]k^{\frac{1}{2}} = k~{\alpha}^N_i(\vec{k}),~~~~\bar{\Delta}_\alpha=2,
\end{align}
also corresponds to the same unitary discrete series representation of $SO(4,1)$, and gives rise to the following late-time states 
\begin{align}
|\tilde{\alpha}^N_i(\vec{k})\rangle&\equiv\tilde{\alpha}^N_i(\vec{k})|0\rangle\\
    &=ik^{\frac{1}{2}}\sum_{\lambda=\pm}\hat{\epsilon}_{i}^{(\lambda)}(\vec{k})|-\vec{k},\lambda\rangle.
    \end{align}

    The split of the photon unitary discrete series representation into a direct sum of irreducible representations with opposite helicity (as shown in (\ref{discrete series in 4D split into two})) is discussed in section \ref{sec:Direct sum of representations in the Hilbert space}. 

\paragraph{Inner product for $\beta_j(\vec{k})$.} 
Let us write the explicit expression for the late-time operator $\beta_j(\vec{k})$ here again for convenience,
\begin{align}
\beta_j(\vec{k})={{\frac{1}{\sqrt{2}}}}~\sum_{\lambda= \pm}  \left[a^{\dagger}_{-\vec{k},\lambda} +a_{\vec{k},\lambda} \right]k^{1/2}~\hat{\epsilon}_{j}^{(\lambda)}(\vec{k}).
\end{align}
Under infinitesimal dilatations (in momentum space), the $\Delta=2$ late-time operator $\beta_j$ transforms as shown in (\ref{dilatation transform of late-time operators}).
Thus, $\beta_j(\vec{k})$ is labeled by $\chi'^+_{0_1}$ (see Table \ref{exceptional table1}), and the corresponding function space  is $C'^+_{0_1}$ \cite{Dobrev:1977qv}. At first sight, defining  an inner product for $\beta_j$ might seem puzzling. The reason is that the intertwiner that acts on the representation space $C'^+_{0_1}$ 
    \begin{align}
       G'^-_{0_1}:C'^+_{0_1}~\to~C'^-_{0_1}, 
    \end{align}
 is expressed (in momentum space) as \cite{Dobrev:1977qv} %
 \footnote{ The full expression from \cite{Dobrev:1977qv} (equation (6.18)), valid for $l=0,1,\dots$ and $\nu=1,2,\dots$ is
\begin{equation}\label{eqn:Gp-int}G'^-_{l_\nu}(\vec{p})=\frac{l!}{\nu!}(d+l-3)!(d+2l+\nu-2)!\left(\frac{2}{p^2}\right)^{\frac{d}{2}+l-1}\sum^{l}_{m=0}\frac{(-1)^m}{(l-m)!(d+l+m-3)!}\Pi^{l+\nu m}(\vec{p}).\end{equation} Note that in this expression the indices have been removed by contracting with null vectors.}%
\begin{align}\label{eqn:betaintert}
     G'^-_{0_1}(\vec{k})=2\left(\frac{2}{k^2}\right)^{\frac{1}{2}}\Pi^{10}(\hat{k}),
\end{align}
where $\Pi^{10}$ projects onto the helicity-0 content \cite{sun2021note},
\begin{align}\label{eqnPi10ij}
    \left[\Pi^{10}(\hat{k})\right]_{ij}=\hat{k}_i\hat{k}_j,~~~\hat{k}_{i}=k_i/k.
\end{align}
We thus find that the intertwiner annihilates $\beta_j$ ,
\begin{align}
  \left[G'^-_{0_1}(\vec{k})\right]_{ij}\beta_j(\vec{k}) =0  ,
\end{align}
because $\beta_j$ includes only the propagating helicities $\pm 1$.
As a consequence, if one attempts to construct an inner product for states created with $\beta_j$ using the intertwiner $G'^-_{0_1}$, then this inner product will be degenerate and will give rise only to zero norms. 

In order to define an invariant inner product for $\beta_j$, let us exploit the isomorphism $D_{0 1} \cong C^{'-}_{01}/F'_{01}$ in (\ref{isomorphisms}), where $\beta_j$ belongs to the representation $D_{0 1}$
(conserved current) and $\alpha_j$ belongs to $C^{'-}_{01}/F'_{01}$ (boundary gauge field with pure-gauge modes modded out)\footnote{For a similar discussion see also \cite{sun2021note}.}. As shown earlier, in practice, the intertwiner $  G'^+_{0_1}(\vec{k})$ acts on the $\Delta=1$ operator through multiplication with a factor of $k$ because the projector $\Pi^{11}$ acts on $\alpha_j$ as the identity. In other words, the operator $\tilde{\alpha}_j=[G_{0_1}^{'+}]_j\,^i\,\alpha_i=k ~\alpha_j$ transforms under dilatations with scaling dimension $\Delta=2$. Motivated by this, using the $\Delta=2$ operator $\beta_j$, we define the $\Delta=1$ operator\footnote{See also \cite{Anninos:2017eib}.}
\begin{align}\label{mapping beta_j to the spaces of a's}
   \beta_j'(\vec{k}) \equiv \frac{1}{k}\beta_{j}(\vec{k}) .
\end{align}
 Using the infinitesimal transformation of creation/annihilation operators under dilatations (\ref{trasnfrmt of annihilation op dilatation}), we can confirm that $\beta_j'(\vec{k})$ transforms as a $\Delta=1$ operator,
\begin{align}
    \delta_{D} \beta'_j(\vec{k}) =& {\frac{1}{\sqrt{2}}}~\sum_{\lambda= \pm}  \left[ \delta_{D}a^{\dagger}_{-\vec{k},\lambda} + \delta_{D}a_{\vec{k},\lambda} \right]k^{-1/2}~\hat{\epsilon}_{j}^{(\lambda)}(\vec{k})\nonumber\\
    =&~\left(k^{i}\frac{\partial}{\partial k^{i}}+2\right)\beta'_j(\vec{k}) .
\end{align}
This means that $\beta'_j(\vec{k})$ belongs to the same representation space with $\alpha_j(\vec{k})$, that is $C^{'-}_{01}$. Now we can define a $SO(4,1)$-invariant inner product for our original operator $\beta_{j}(\vec{k})$ as follows:
\begin{align}\label{eq:inner product realization 2}
\langle \beta|\beta \rangle \equiv\frac{1}{\Omega}\left(\beta',G^{'+}_{01}\beta'\right)& \equiv\frac{1}{\Omega}\int \frac{d^3k}{(2\pi)^3}~\frac{1}{k^2}\langle\beta_i(\vec{k})|[G_{01}^{'+}]_{ij}|{\beta}_j(\vec{k})\rangle ,
\end{align}
where $\frac{1}{\Omega}\left(\cdot , \cdot\right)$ is the densitized inner product defined in (\ref{eqn:densitizedinner_alpha}). The result is
    \begin{align}
\langle \beta|\beta \rangle = {\frac{1}{2}}.
\end{align}

As in the case of $\alpha_j$, we can define a normalized operator 
\begin{align}
\label{betaN}
{\beta}^N_i(\vec{k})=\sum_{\lambda=\pm}\hat{\epsilon}_{i}^{(\lambda)}(\vec{k})\left[\admk+\ak\right]k^{\frac{1}{2}},~~~~{\Delta}_\beta=2.
\end{align}
States created by ${\beta}^N_i(\vec{k})$ 
\begin{align}
|\beta^N_i(\vec{k})\rangle&\equiv {\beta}^N_i(\vec{k})|0\rangle,
    \end{align}
also furnish the photon unitary discrete series representation of $SO(4,1)$, i.e. $D_{01}$ (\ref{discrete series in 4D split into two}). The split of this representation into a direct sum of irreducible representations with opposite helicity will be demonstrated in section \ref{sec:Direct sum of representations in the Hilbert space}.



\section{The de Sitter invariance of the (anti-)self-duality condition on the field strength tensor}
\label{sec:field strength and anti-self-duality}

As a `useful tool' in order to demonstrate that the electromagnetic field contains two representations with opposite helicity, let us introduce the gauge-invariant field strength tensor in the bulk
\begin{align}
    F_{\mu \nu} = \partial_\mu  A_{\nu}-\partial_{\nu}A_{\mu}.
\end{align}
 The split of the photon discrete series representation of $SO(4,1)$ into a direct sum of representations with opposite helicity will be demonstrated in section \ref{sec:Direct sum of representations in the Hilbert space}. In the present section, we discuss some properties of the field strength, as well as its self-dual and anti-self-dual parts, which will be useful in our discussion in section \ref{sec:Direct sum of representations in the Hilbert space}.

It is useful to work with the field strength because the two discrete series (discrete series $+$ and $-$) correspond to massless particles of opposite helicities on dS spacetime%
\footnote{This is well-understood in global de Sitter spacetime, for all massless and partially spinning fields \cite{Silva_note, Higuchi:STSHS, Letsios:2022tsq, Letsios:2023awz, Letsios:2023qzq}).},~%
while the field strength modes can be also split into two parts corresponding to  photons of two opposite helicities. 
In particular, using the Bunch-Davies modes for the gauge potential, we can readily obtain the field strength modes as
\begin{align}
  F^{(\lambda,\vec{k})}_{\mu \nu}(\vec{x},\eta)=2~\partial_{[\mu}  A^{(\lambda,\vec{k})}_{\nu]}(\vec{x},\eta) \equiv  \partial_{\mu}  A^{(\lambda,\vec{k})}_{\nu}(\vec{x},\eta) - \partial_{\nu}  A^{(\lambda,\vec{k})}_{\mu}(\vec{x},\eta) ,
\end{align}
with components  given by 
\begin{align}\label{field strength modes explct componts}
   & F^{(\lambda,\vec{k})}_{\eta j}(\vec{x},\eta)= - \frac{1}{(2\pi)^{3/2}}\sqrt{k/2}~e^{ik|\eta|+i\vec{k}\cdot  \vec{x}}~\hat{\epsilon}^{(\lambda)}_{j}(\vec{k}), \nonumber\\
   &F^{(\lambda,\vec{k})}_{i j}(\vec{x},\eta)= \frac{1}{(2\pi)^{3/2}} \sqrt{1/(2k)}~e^{ik|\eta|+i\vec{k}\cdot  \vec{x}}~(k_{i}~\hat{\epsilon}^{(\lambda)}_{j}(\vec{k})-  k_{j}~\hat{\epsilon}^{(\lambda)}_{i}(\vec{k})).
\end{align}
As we will discuss shortly, the modes $F^{(+,\vec{k})}_{\mu \nu}$  ($F^{(-,\vec{k})*}_{\mu \nu}$) are  positive (negative) frequency  modes that are self-dual and describe photons of fixed helicity $+ 1$ ($-1$), as in the case of Minkowski spacetime. Similarly, $F^{(-,\vec{k})}_{\mu \nu}$  ($F^{(+,\vec{k})*}_{\mu \nu}$) are  positive (negative) frequency modes  that are anti-self-dual, corresponding to photons of fixed helicity $- 1$ ($+1$).

\subsection{Mode expansion in the bulk}
The quantum field strength is expanded in modes by simply using the mode expansion of the gauge potential, as
\begin{align}
    F_{\mu \nu}(\vec{x},\eta) = \sum_{\lambda=\pm}\int \frac{d^{3}k}{(2   \pi)^{3/2}} \left( a_{\vec{k},\lambda} ~F^{(\lambda,\vec{k})}_{\mu \nu}(\vec{x},\eta)+ a^{\dagger}_{\vec{k},\lambda} ~F^{(\lambda,\vec{k})}_{\mu \nu}(\vec{x},\eta)^{*} \right).
\end{align}
More explicitly,
\begin{align}
 F_{\eta j}(\vec{x},\eta) = \sum_{\lambda=\pm}\int \frac{d^{3}k}{(2   \pi)^{3}} e^{i \vec{k}\cdot  \vec{x}}\left( -a_{\vec{k},\lambda} \sqrt{\frac{k}{2}}~e^{ik|\eta|}~\hat{\epsilon}^{(\lambda)}_{j}(\vec{k})-a^{\dagger}_{-\vec{k},\lambda} \sqrt{\frac{k}{2}}~e^{-ik|\eta|}~\hat{\epsilon}^{(\lambda)}_{j}(\vec{k})\right).
\end{align}
\begin{align}
 F_{i j}(\vec{x},\eta) = \sum_{\lambda=\pm}\int \frac{d^{3}k}{(2   \pi)^{3}} e^{i \vec{k}\cdot  \vec{x}}\left( a_{\vec{k},\lambda} \sqrt{\frac{1}{2k}}~e^{ik|\eta|}~2~k_{[i}~\hat{\epsilon}^{(\lambda)}_{j]}(\vec{k})-a^{\dagger}_{-\vec{k},\lambda} \sqrt{\frac{1}{2k}}~e^{-ik|\eta|}~2~k_{[i}~\hat{\epsilon}^{(\lambda)}_{j]}(\vec{k})\right).
\end{align}

Let us now split the field strength into two parts, $F_{\mu \nu}=F^{\text{(SD)}}_{\mu \nu}+F^{\text{(ASD)}}_{\mu \nu}$, where  
\begin{align}\label{def:field strength mode exns covariant}
    F^{\text{(SD)}}_{\mu \nu}(\vec{x},\eta) &= \int \frac{d^{3}k}{(2   \pi)^{3/2}} \left( a_{\vec{k},+} ~F^{(+,\vec{k})}_{\mu \nu}(\vec{x},\eta)+ a^{\dagger}_{\vec{k},-} ~F^{(-,\vec{k})}_{\mu \nu}(\vec{x},\eta)^{*} \right), \nonumber \\
    F^{\text{(ASD)}}_{\mu \nu}(\vec{x},\eta) &= \int \frac{d^{3}k}{(2   \pi)^{3/2}} \left( a_{\vec{k},-} ~F^{(-,\vec{k})}_{\mu \nu}(\vec{x},\eta)+ a^{\dagger}_{\vec{k},+} ~F^{(+,\vec{k})}_{\mu \nu}(\vec{x},\eta)^{*} \right).
\end{align}
The labels {$(\text{SD})$} and {$(\text{ASD})$} stand for self-dual and anti-self-dual, respectively, and they will be discussed further in the following section. In particular, we have
\begin{align}\label{field strngth +- component eta j}
 F^{\text{(ASD/SD)}}_{\eta j}(\vec{x},\eta) = \int \frac{d^{3}k}{(2   \pi)^{3}} e^{i \vec{k}\cdot  \vec{x}}\left( -a_{\vec{k},\mp} \sqrt{\frac{k}{2}}~e^{ik|\eta|}~\hat{\epsilon}^{(\mp)}_{j}(\vec{k})-a^{\dagger}_{-\vec{k},\pm} \sqrt{\frac{k}{2}}~e^{-ik|\eta|}~\hat{\epsilon}^{(\pm)}_{j}(\vec{k})\right),
\end{align}
\begin{align}\label{field strngth +- component i j}
 F^{\text{(ASD/SD)}}_{i j}(\vec{x},\eta) = \int \frac{d^{3}k}{(2   \pi)^{3}} e^{i \vec{k}\cdot  \vec{x}}\left( a_{\vec{k},\mp} \sqrt{\frac{1}{2k}}~e^{ik|\eta|}~2~k_{[i}~\hat{\epsilon}^{(\mp)}_{j]}(\vec{k})-a^{\dagger}_{-\vec{k},\pm} \sqrt{\frac{1}{2k}}~e^{-ik|\eta|}~2~k_{[i}~\hat{\epsilon}^{(\pm)}_{j]}(\vec{k})\right).
\end{align}
Below we will show that the aforementioned split of the field strength into $F^{\text{(SD)}}_{\mu \nu}$ and $F^{\text{(ASD)}}_{\mu \nu}$ is  invariant under infinitesimal dS transformations. This means that the self-dual sector does not mix with the anti-self-dual sector (and vice versa) under any $so(4,1)$ transformation. 


\subsection{Anti-self-duality and self-duality} \label{subsect:duality discussions}
Let $\epsilon_{\mu \nu \rho \sigma}$ be the totally anti-symmetric Levi-Civita tensor with $\epsilon_{\eta x y z}=\sqrt{-g}=H^{-4}|\eta|^{-4}$. The dual field strength is defined as
\begin{align}\label{def:duality transfrmn}
    \widetilde{F}_{\mu \nu}=\frac{1}{2}\epsilon_{\mu \nu}\,^{\rho \sigma} \, F_{\rho \sigma}.
\end{align}
The self-dual ($F^{\text{(SD)}}_{\mu \nu}$) and anti-self-dual ($F^{\text{(ASD)}}_{\mu \nu}$) field strengths are those field strengths that satisfy
\begin{align}\label{(anti-)selfdty condtion}
    \widetilde{F}^{\text{(SD)}}_{\mu \nu}=+ \,i\, F_{\mu \nu}^{\text{(SD)}},~ ~~~\widetilde{F}^{\text{(ASD)}}_{\mu \nu}=- \,i\, F_{\mu \nu}^{\text{(ASD)}}.
\end{align}
Any field strength can be split into these two parts as $F_{\mu \nu}=F^{\text{(SD)}}_{\mu \nu} + F^{\text{(ASD)}}_{\mu \nu}$. For real $F_{\mu \nu}$, the self-dual and anti-self-dual parts are complex conjugates of each other,

Let us now verify that our mode expansion (\ref{def:field strength mode exns covariant}) for the field strength in terms of Bunch-Davies modes is consistent with the split into self-dual and anti-self-dual parts.
Using the explicit expressions for the field-strength Bunch-Davies modes (\ref{field strength modes explct componts}), we find by direct calculation that helicity $\lambda =\pm$   `controls' their (anti-)self-duality, in the sense that 
\begin{align}
    &\widetilde{F}^{(\lambda,\vec{k})}_{\mu \nu}= i\, \lambda \,  F^{(\lambda,\vec{k})}_{\mu \nu},~~~~~~~~~\widetilde{F}^{(\lambda,-\vec{k})
    }_{\mu \nu}= i\, \lambda \,  F^{(\lambda,-\vec{k})}_{\mu \nu} ,\nonumber\\
    &\widetilde{F}^{(\lambda,-\vec{k})*}_{\mu \nu}= -i\, \lambda \,  F^{(\lambda,-\vec{k})*}_{\mu \nu},~~~\widetilde{F}^{(\lambda,\vec{k})
    *}_{\mu \nu}= -i\, \lambda \,  F^{(\lambda,\vec{k})*}_{\mu  \nu}.
\end{align}
Then, it follows immediately that the mode expansion of the field strengths $F^{\text{(SD)}}_{\mu \nu}$  and $F^{\text{(ASD)}}_{\mu \nu}$ in (\ref{def:field strength mode exns covariant}) is consistent with the self-duality and anti-self-duality conditions (\ref{(anti-)selfdty condtion}), respectively. We have thus verified that our expressions for $F^{\text{(SD)}}_{\mu \nu}$  and $F^{\text{(ASD)}}_{\mu \nu}$ in eq. (\ref{def:field strength mode exns covariant}) coincide with the self-dual and anti-self-dual parts, respectively, of the field strength.

\subsection{dS invariance of (anti-)self-dual sectors}
\label{subsec:dS invariance and a direct sum on the space of classical solutions}
Let us now show that the (anti-)self-duality properties of the field strength are preserved under any infinitesimal dS transformation. We first observe that $\epsilon_{\mu \nu}\,^{\rho \sigma}=g^{\rho \rho'} g^{\sigma \sigma'} \epsilon_{\mu \nu \rho'  \sigma'}$ is constant ($\partial_{\alpha}\epsilon_{\mu \nu}\,^{\rho \sigma}=0$) with $\epsilon_{\eta x}\,^{yz}=1$. Then, it is easy to show that $\epsilon_{\mu \nu}\,^{\rho \sigma}$ is invariant under any infinitesimal dS transformations. In particular, the infinitesimal change of $\epsilon_{\mu \nu}\,^{\rho \sigma}$ under the dS transformation generated by any Killing vector $\xi$ is zero, as
\begin{align}
   \mathcal{L}_{\xi} \epsilon_{\mu \nu}\,^{\rho \sigma}=& \partial_{\mu}\xi^{\lambda}~\epsilon_{\lambda \nu}\,^{\rho \sigma}+\partial_{\nu}\xi^{\lambda}~\epsilon_{\mu \lambda }\,^{\rho \sigma}-\partial_{\lambda}\xi^{\rho}~\epsilon_{\mu \nu}\,^{\lambda \sigma} - \partial_{\lambda}\xi^{\sigma}~\epsilon_{\mu \nu}\,^{\rho \lambda} \nonumber \\
   =&0.
\end{align}
This can be easily checked by direct calculation for all ten Killing vectors. Then, from eqs. (\ref{def:duality transfrmn}) and (\ref{(anti-)selfdty condtion}), we find 
\begin{align}\label{[duality,ds]Fmn=0}
    \frac{1}{2}\epsilon_{\mu \nu}\,^{\rho \sigma} \, \mathcal{L}_{\xi}F^{\text{(SD)}}_{\rho \sigma}=+ i\,\mathcal{L}_{\xi}{F}^{\text{(SD)}}_{\mu \nu},~~~\frac{1}{2}\epsilon_{\mu \nu}\,^{\rho \sigma} \, \mathcal{L}_{\xi}F^{\text{(ASD)}}_{\rho \sigma}=- i\,\mathcal{L}_{\xi}{F}^{\text{(ASD)}}_{\mu \nu}
\end{align}
for any Killing vector. According to this equation, if we consider a(n) (anti-)self-dual field strength $F^{\text{(ASD/SD)}}_{\mu \nu}$ and we act on it with any infinitesimal dS transformation, then the transformed field strength $\mathcal{L}_{\xi}F^{\text{(ASD/SD)}}_{\mu \nu}$ is also (anti-)self-dual. In other words, self-dual and anti-self-dual field strengths never mix with each other under infinitesimal dS transformations. This means that, on the field-strength solution space of Maxwell equations, the Lie derivatives with respect to dS Killing vectors generate a direct sum of $so(4,1)$ representations.


 \section{Identifying the direct sum  `discrete series $+$ $\oplus$ discrete series $-$' at late times}
 \label{sec:Direct sum of representations in the Hilbert space}
 For convenience, let us give here again the form of our late-time operators introduced in \eqref{eqn:introducing alpha beta k}, 
 \begin{subequations}
\label{eqn:introducing alpha beta k_rewrite}
    \begin{align}
\label{eqn:introducingalphak_rw}\alpha_j(\vec{k})&=i{\frac{1}{\sqrt{2}}}\sum_{\lambda=\pm}\left[\admk-\ak\right]k^{-\frac{1}{2}}~\hat{\epsilon}_{j}^{(\lambda)}(\vec{k}),\\
\label{eqn:introducingbetak_rw}\beta_j(\vec{k})&={\frac{1}{\sqrt{2}}}\sum_{\lambda=\pm}\left[\admk+\ak\right]k^{\frac{1}{2}}~\hat{\epsilon}_{j}^{(\lambda)}(\vec{k}).
    \end{align}
\end{subequations}
We discussed the normalization of these operators with respect to a CFT-inspired inner product in section \ref{subsec:The discrete series inner product for beta}. We demonstrated that $\alpha_j(\vec{k})$ and $\beta_j(\vec{k})$ create normalizable states with positive norm, furnishing the unitary discrete series representations of $SO(4,1)$ associated with the photon. 

Our goal in this section is to determine the decomposition of the late-time operators into a `{$+$} part' and a `{$-$} part', as 
\begin{align}
    \alpha_j(\vec{k})&=\alpha_j^{(+)}(\vec{k})+\alpha_j^{(-)}(\vec{k}),\\
    \beta_j(\vec{k})&=\beta_j^{(+)}(\vec{k})+\beta_j^{(-)}(\vec{k}).
\end{align}
Each part will be shown to create states (when acting on the Bunch-Davies vacuum) that separately form irreducible representations of $SO(4,1)$. In particular, we have a direct sum of representations with opposite helicity. That is, the direct sum of unitary representations: `discrete series $+$ $\bigoplus $ discrete series $-$' (\ref{discrete series in 4D split into two}).

  The explicit expressions for the $+$ and $-$ parts of the late-time operators are
\begin{subequations} \label{eqn:alpha beta lambda}
    \begin{align}
 \alpha_j^{(\lambda)}(\vec{k})&=i{{\frac{1}{\sqrt{2}}}}\left[\hat{\epsilon}_{j}^{(\lambda)}(\vec{k})\admk-\hat{\epsilon}_{j}^{(-\lambda)}(\vec{k})a_{\vec{k},-\lambda}\right]k^{-1/2},\\
   \beta_j^{(\lambda)}(\vec{k})&={\frac{1}{\sqrt{2}}}\left[\hat{\epsilon}_{j}^{(\lambda)}(\vec{k})\admk+\hat{\epsilon}_{j}^{(-\lambda)}(\vec{k})a_{\vec{k},-\lambda}\right]k^{1/2}, ~~\lambda = \pm.
    \end{align}
\end{subequations}
 Note that
\begin{align} 
\alpha_j^{(-\lambda)}(\vec{k}) =  \alpha_j^{(\lambda)}(-\vec{k})^{\dagger},~~~\beta_j^{(-\lambda)}(\vec{k}) =  \beta_j^{(\lambda)}(-\vec{k})^{\dagger}  .
\end{align} 
Before proceeding, let us develop some intuition concerning why the $+$ and $-$ parts above do not mix under dS transformations. In particular, let us show that $\alpha_j^{(-)}(\vec{k})$ and $\beta_j^{(-)}(\vec{k})$ are the late-time operators corresponding to the gauge potential of the self-dual field strength, and similarly,  $\alpha_j^{(+)}(\vec{k})$ and $\beta_j^{(+)}(\vec{k})$ are the late-time operators  corresponding to the  gauge potential of the anti-self-dual field strength. These will become clear in eqs. (\ref{field strength late time op A}), (\ref{field strength late time op B}), (\ref{field strength late time op A and B -- relating SD and ASD}) below. Since self-dual and anti-self-dual field strengths  do not mix with each other under $so(4,1)$ transformations (see Section \ref{subsec:dS invariance and a direct sum on the space of classical solutions}), it is natural that their corresponding gauge potentials do not mix under $so(4,1)$ either (this will be proved explicitly in Subsection \ref{subsec: negative and pos do NOT mix late time proof}).
\subsection{Late-time operators for the (anti-)self-dual field strength}

Let us take the late-time limit of the self-dual field strength 
\begin{align}
    \lim_{|\eta|\to0} F^{(\text{SD})}_{\mu \nu}(\vec{x},\eta)= \int \frac{d^{3}k}{(2   \pi)^{3/2}} \left( a_{\vec{k},+} ~\lim_{|\eta|\to0}F^{(+,\vec{k})}_{\mu \nu}(\vec{x},\eta)+ a^{\dagger}_{\vec{k},-} ~\lim_{|\eta|\to0}F^{(-,\vec{k})}_{\mu \nu}(\vec{x},\eta)^{*} \right).
\end{align}
We find 
    \begin{align}
\label{eqn:F_mu nu latetime SD}        \lim_{|\eta|\to0} F^{(\text{SD})}_{\mu \nu}(\vec{x},\eta)&=\int\frac{d^3k}{(2\pi)^3}e^{i\vec{k}\cdot\vec{x}}\Bigg[|\eta|^0 \, \mathbb{A}^{(-)}_{\mu \nu}(\vec{k})+|\eta|\, \mathbb{  B}^{(-)}_{\mu \nu}(\vec{k})\Bigg]+\mathcal{O}(\eta^2), 
\end{align}
where the late-time operators $\mathbb{A}^{(-)}_{\mu \nu}(\vec{k}), \mathbb{B}^{(-)}_{\mu \nu}(\vec{k})$ are anti-symmetric tensors. Their components are given in terms of the gauge-potential late-time operators, as
\begin{align}\label{field strength late time op A}
   \mathbb{A}^{(-)}_{\eta j}(\vec{k}) = -\beta^{(-)}_{j}(\vec{k}),~~\mathbb{A}^{(-)}_{m j}(\vec{k}) =2\, i~ k_{[m}~\alpha^{(-)}_{j]}(\vec{k}),
\end{align}
\begin{align}\label{field strength late time op B}
   \mathbb{B}^{(-)}_{\eta j}(\vec{k}) = k^2~\alpha^{(-)}_{j}(\vec{k}),~~\mathbb{B}^{(-)}_{m j}(\vec{k}) =2\, i~ k_{[m}~\beta^{(-)}_{j]}(\vec{k}).
\end{align}
One can readily check that each of the two late-time operators, $\mathbb{A}^{(-)}_{\mu \nu}(\vec{k})$ and  $\mathbb{B}^{(-)}_{\mu \nu}(\vec{k})$, is self-dual.

The late-time limit of the anti-self-dual field strength can be found similarly
   \begin{align}
\label{eqn:F_mu nu latetime anti-SD}        \lim_{|\eta|\to0} F^{(\text{ASD})}_{\mu \nu}(\vec{x},\eta)&=\int\frac{d^3k}{(2\pi)^3}e^{i\vec{k}\cdot\vec{x}}\Bigg[|\eta|^0 \, \mathbb{A}^{(+)}_{\mu \nu}(\vec{k})+|\eta|\, \mathbb{  B}^{(+)}_{\mu \nu}(\vec{k})\Bigg]+\mathcal{O}(\eta^2), 
\end{align}
where, now, the late-time operators are anti-self-dual anti-symmetric tensors.
Equation \eqref{eqn:F_mu nu latetime anti-SD} is related to (\ref{eqn:F_mu nu latetime SD}) via hermitian conjugation, which means 
\begin{align}\label{field strength late time op A and B -- relating SD and ASD}
\mathbb{A}^{(-)}_{\mu \nu}(-\vec{k})^{\dagger} = \mathbb{A}^{(+)}_{\mu \nu}(\vec{k}),~~\mathbb{B}^{(-)}_{\mu \nu}(-\vec{k})^{\dagger} = \mathbb{B}^{(+)}_{\mu \nu}(\vec{k}).\end{align}
This is straightforward to check. Recall that $\dagger$ refers to hermitian conjugation with respect to the bulk QFT inner product.

 We have thus verified that  $\alpha_j^{(-)}(\vec{k})$ and $\beta_j^{(-)}(\vec{k})$ are the late-time gauge-potential operators corresponding to the self-dual field strength, while $\alpha_j^{(+)}(\vec{k})$ and $\beta_j^{(+)}(\vec{k})$ correspond to the anti-self-dual field strength.

\subsection{Negative-helicity and positive-helicity states do not mix under dS transformations}\label{subsec: negative and pos do NOT mix late time proof}
In this Subsection, we will show that photon states of fixed helicity form a direct sum of representations, corresponding to the two helicities $+1$ and $-1$. We will first discuss the transformation of fixed-helicity states by taking advantage of properties of  the bulk field strength,
Then, we will demonstrate that late-time operators with fixed helicity labels, $\alpha^{(+)}_j$ and $\alpha^{(-)}_j$, do not mix with each other under $so(4,1)$, i.e. they furnish a direct sum of unitary discrete series representations. We will similarly show that  $\beta^{(+)}_j$ and $\beta^{(-)}_j$ furnish a direct sum of unitary discrete series representations of $so(4,1)$. The discussion concerning the CFT-inspired norm in each  summand in the direct sum of representations is identical with the discussion in section \ref{subsec:The discrete series inner product for beta}. That is, the CFT-inspired inner product for $\alpha^{(+)}_j$ ($\beta^{(+)}_j$) and $\alpha^{(-)}_j$ ($\beta^{(-)}_j$) is the same with the one used for $\alpha_j$ ($\beta_j$) in section \ref{subsec:The discrete series inner product for beta}.

Consider the bulk self-dual and anti-self-dual field strength operators (\ref{def:field strength mode exns covariant}). For convenience, we take them to act on the Bunch-Davies vacuum
\begin{align}
    F^{\text{(SD)}}_{\mu \nu}(\vec{x},\eta) |0 \rangle&= \int \frac{d^{3}k}{(2   \pi)^{3/2}}   ~F^{(-,\vec{k})}_{\mu \nu}(\vec{x},\eta)^{*}~~a^{\dagger}_{\vec{k},-} |0 \rangle , \\
     F^{\text{(ASD)}}_{\mu \nu}(\vec{x},\eta) |0 \rangle&= \int \frac{d^{3}k}{(2   \pi)^{3/2}}   ~F^{(+,\vec{k})}_{\mu \nu}(\vec{x},\eta)^{*}~~a^{\dagger}_{\vec{k},+} |0 \rangle .
\end{align}
Recall, from section \ref{sec:field strength and anti-self-duality}, that the (anti-)self-dual field  strength maintains its (anti-)self-duality under $so(4,1)$ transformations. We can thus write the following expression for the infinitesimal dS transformations of the (anti-)self-dual modes 
\begin{align}\label{Lie deriv of classical field-strength modes}
&\mathcal{L}_{\xi}F^{(\lambda,\vec{k})}_{\mu \nu}(\vec{x},\eta)^{*} = \int d^3q~m^{(\lambda)}_{\xi}(\vec{k}|\vec{q})~F^{(\lambda,\vec{q})}_{\mu \nu}(\vec{x},\eta)^{*},~~\lambda \in \{ +,-\} ,
\end{align}
where the complex coefficients $m^{(\lambda)}_{\xi}(\vec{k}|\vec{q})$  satisfy $m^{(\lambda)}_{\xi}(\vec{k}|\vec{q})^{*} = - m^{(\lambda)}_{\xi}(\vec{q} | \vec{k}) $ due to the anti-hermiticity of $\mathcal{L}_\xi$ with respect to the Klein-Gordon inner product,
for any Killing vector $\xi$ -- see section \ref{sec:Warm up: The QFT Hilbert space}.~
{Also note that the Lie derivative $\mathcal{L}_{\xi}$ along any Killing vector $\xi^\mu$ commutes with covariant derivatives on de Sitter spacetime}%
\footnote{One can readily show that the covariant derivative of a tensor field in de Sitter spacetime commutes with the Lie derivative $\mathcal{L}_{\xi}$ where $\xi^\mu$ is a Killing vector. For example, in the case of the spin-1 field we find by direct calculation $[\mathcal{L}_\xi , \nabla_\mu ]A_\nu=0$. To perform the calculation, we computed a commutator of two covariant derivatives (the second covariant derivatives appears by expanding $\mathcal{L}_\xi$), using the well-known formula for the Riemann tensor of $dS_4$, $R_{\mu \nu \rho \sigma} =H^2( g_{\mu \rho} ~g_{\nu \sigma} -(\mu \leftrightarrow \nu))$. Moreover, we made use of the well-known formula for two covariant derivatives acting on a Killing vector $\nabla_\mu \nabla_\nu \xi_\rho = R_{\rho \nu \mu \sigma} \xi^\sigma$.}%
, and thus, 
\begin{align}
\mathcal{L}_{\xi}F^{(\lambda,\vec{k})}_{\mu \nu}(\vec{x},\eta)^{*} = 2 ~\partial_{[\mu}~\mathcal{L}_{\xi}A^{(\lambda,\vec{k})}_{ \nu]}(\vec{x},\eta)^{*},
\end{align}
which means that the field-strength modes `inherit' their transformation properties from the gauge-potential Bunch-Davies modes $A^{(\lambda,\vec{k})}_{ \nu}(\vec{x},\eta)$, with the advantage that any pure-gauge contribution stemming from the transformation of  $A^{(\lambda,\vec{k})}_{ \nu}(\vec{x},\eta)$ drops out.
Equation (\ref{Lie deriv of classical field-strength modes}) means that (anti-)self-dual field strength modes transform among themselves.\footnote{We do not need explicit expressions for the coefficients $m^{(\lambda)}_{\xi}(\vec{k}|\vec{q})$ for our analysis. However, note that, for the case of dilatations and translations,   explicit expressions can be deduced from eqs. (\ref{trasnfrmt of annihilation op dilatation}) and (\ref{trasnfrmt of annihilation op translation}), respectively.}

Now, since the infinitesimal transformation $\delta_\xi  F^{\text{(SD)}}_{\mu \nu} \equiv - \mathcal{L}_\xi  F^{\text{(SD)}}_{\mu \nu}$ (and similarly $\delta_\xi  F^{\text{(ASD)}}_{\mu \nu}$) can be attributed to either the classical modes or to the creation/annihilation operators, we find
\begin{align}\label{dS transformation of (A)SD bulk F_mn}
    \delta_\xi F^{\text{(SD)}}_{\mu \nu}(\vec{x},\eta) |0 \rangle&= \int \frac{d^{3}k}{(2   \pi)^{3/2}}   ~F^{(-,\vec{k})}_{\mu \nu}(\vec{x},\eta)^{*}~~\delta_\xi a^{\dagger}_{\vec{k},-} |0 \rangle , \nonumber \\
     \delta_\xi F^{\text{(ASD)}}_{\mu \nu}(\vec{x},\eta) |0 \rangle&= \int \frac{d^{3}k}{(2   \pi)^{3/2}}   ~F^{(+,\vec{k})}_{\mu \nu}(\vec{x},\eta)^{*}~~\delta_\xi a^{\dagger}_{\vec{k},+} |0 \rangle ,
\end{align}
where creation operators of fixed helicity transform among themselves, as
\begin{align} \label{dS transform of creation op generic and helicity preservation}
&\delta_{\xi}a_{\vec{k},\lambda}^{\dagger}= \int d^3q~m^{(\lambda)}_{\xi}(\vec{k}|\vec{q})^{*}~a_{\vec{q},\lambda}^{\dagger} ,~~\lambda \in \{ +,-\}.
\end{align}
This equation reveals that the two families of single-particle states with fixed helicity, $a_{\vec{k},+}^{\dagger}|0\rangle$ and $a_{\vec{k},-}^{\dagger}|0\rangle$, form a direct sum of $so(4,1)$ representations. Indeed, the $so(4,1)$ charges (\ref{hermitian quantum dS charges}) act on single-particle states as
\begin{align}
Q(\xi)~a_{\vec{k},\lambda}^{\dagger}|0\rangle&= -\left[a_{\vec{k},\lambda}^{\dagger},Q(\xi) \right]|0\rangle=i\delta_{\xi}a_{\vec{k},\lambda}^{\dagger}|0\rangle  \nonumber\\
&= i\int d^3q~m^{(\lambda)}_{\xi}(\vec{k}|\vec{q})^{*}~a_{\vec{q},\lambda}^{\dagger}|0\rangle,
\end{align}
where helicity $\lambda \in \{+,-\}$ is preserved. This explains the split of the photon single-particle Hilbert space into a direct sum of $so(4,1)$ representations with opposite helicity.\footnote{From eqs. (\ref{trasnfrmt of annihilation op dilatation}) and (\ref{trasnfrmt of annihilation op translation}), we saw that helicity is preserved under dilatations and translations. However, the discussion we give here demonstrates that helicity is preserved under all $so(4,1)$ transformations.}

For completeness, let us discuss how this split manifests itself at the level of states created by the late-time operators. Taking the late-time limit of (\ref{dS transformation of (A)SD bulk F_mn}), and focusing on the spatial components, we find
 \begin{align}        \lim_{|\eta|\to0} \delta_\xi F^{(\text{SD})}_{jl}(\vec{x},\eta)|0\rangle&=\int\frac{d^3k}{(2\pi)^3}e^{i\vec{k}\cdot\vec{x}}\Bigg[|\eta|^0 \, ~\delta_\xi \mathbb{A}^{(-)}_{jl}(\vec{k})|0\rangle+|\eta|\, ~\delta_\xi \mathbb{  B}^{(-)}_{jl}(\vec{k})|0\rangle\Bigg], \nonumber \\
       \lim_{|\eta|\to0} \delta_\xi F^{(\text{ASD})}_{jl}(\vec{x},\eta)|0\rangle&=\int\frac{d^3k}{(2\pi)^3}e^{i\vec{k}\cdot\vec{x}}\Bigg[|\eta|^0 \, ~\delta_\xi \mathbb{A}^{(+)}_{jl}(\vec{k})|0\rangle+|\eta|\, ~\delta_\xi \mathbb{  B}^{(+)}_{jl}(\vec{k})|0\rangle\Bigg],
\end{align}
where we also used (\ref{eqn:F_mu nu latetime SD}) and (\ref{eqn:F_mu nu latetime anti-SD}).  From eqs. (\ref{field strength late time op A}), (\ref{field strength late time op B}) and (\ref{field strength late time op A and B -- relating SD and ASD}), the transformed gauge-potential late-time operators are expressed in terms of  $\delta_\xi \mathbb{  A}^{(\pm)}_{jl}$ and $\delta_\xi\mathbb{  B}^{(\pm)}_{jl}$ as 
\begin{align}
 \delta_\xi\alpha^{(\pm)}_{l}(\vec{k})= -i \frac{k^j}{k^2}  \delta_\xi\mathbb{A}^{(\pm)}_{jl}(\vec{k}) ,~~~~\delta_\xi\beta^{(\pm)}_{l}(\vec{k})= -i \frac{k^j}{k^2}  \delta_\xi\mathbb{B}^{(\pm)}_{jl}(\vec{k}) ,
\end{align}
where $\delta_\xi$ acts on creation/annihilation operators as in (\ref{dS transform of creation op generic and helicity preservation}).
Then, using the explicit expressions for the late-time operators (\ref{eqn:alpha beta lambda}), we find 
    \begin{align}
&Q(\xi)~|\alpha_j^{(\lambda)}(\vec{k})  \rangle=i\,\delta_\xi \alpha_j^{(\lambda)}(\vec{k}) |0 \rangle=i\, {{\frac{1}{\sqrt{2}}}}k^{-1/2}~\hat{\epsilon}_{j}^{(\lambda)}(\vec{k})~[Q(\xi),\admk] ~|0 \rangle ,\\
 &Q(\xi)~|\beta_j^{(\lambda)}(\vec{k})  \rangle=  i \delta_\xi\beta_j^{(\lambda)}(\vec{k}) |0 \rangle={{\frac{1}{\sqrt{2}}}}\,k^{1/2}~\hat{\epsilon}_{j}^{(\lambda)}(\vec{k})~[Q(\xi),\admk] ~|0 \rangle, ~~\lambda = \pm,
    \end{align}
    where $[Q(\xi),\admk] $ is found from eq. (\ref{dS transform of creation op generic and helicity preservation}).
One readily observes that the transformed states $Q(\xi)~|\alpha_j^{(\lambda)}(\vec{k})  \rangle$ and $Q(\xi)~|\beta_j^{(\lambda)}(\vec{k})  \rangle$ are transverse. Also, it is useful to note that the late-time states with fixed helicity labels are eigenstates of the helicity operator (see eq. (\ref{eq:pol vec})), as
\begin{align}
  & h  |\alpha_j^{(\lambda)}(\vec{k}) \rangle=-i
\,\lambda |\alpha_j^{(\lambda)}(\vec{k}) \rangle,~~h  |\beta_j^{(\lambda)}(\vec{k}) \rangle=-i
\,\lambda |\beta_j^{(\lambda)}(\vec{k}) \rangle,
\end{align}
where the helicity operator $h$ acts on vectors $u_l$ on $\mathbb{R}^3$ as
\begin{align}\label{def: helicity operator}
    (h u)_l \equiv \frac{k^i}{k} \epsilon_{lij}u^j.
\end{align}
We then observe that dS charges $Q(\xi)$ do not change the  helicity of late-time states, as
\begin{align}
  h ~ Q(\xi)|\alpha_j^{(\lambda)}(\vec{k}) \rangle=-i
\,\lambda ~Q(\xi)|\alpha_j^{(\lambda)}(\vec{k}) \rangle,~~h ~Q(\xi) |\beta_j^{(\lambda)}(\vec{k}) \rangle=-i
\,\lambda ~Q(\xi)|\beta_j^{(\lambda)}(\vec{k}) \rangle.
\end{align}
These equations can be readily verified as $Q(\xi)$ acts on the creation and annihilation operators inside the late-time operators through the commutator
\begin{align}
   Q(\xi)|\alpha_j^{(\lambda)}(\vec{k}) \rangle =  \left[Q(\xi),\alpha_j^{(\lambda)}(\vec{k})\right] |0\rangle,~~ ~~~  Q(\xi)|\beta_j^{(\lambda)}(\vec{k}) \rangle =  \left[Q(\xi),\beta_j^{(\lambda)}(\vec{k})\right] |0\rangle,
\end{align}
while $h$ acts on the polarization vectors.
Thus, the two sets of states $\{ |\alpha_j^{(+)}(\vec{k}) \rangle\}$ and $\{ |\alpha_j^{(-)}(\vec{k}) \rangle\}$, which are eigenstates of the helicity operator, do not mix under $so(4,1)$ transformations generated by any dS charge $Q(\xi)$. In other words, the two sets of fixed-helicity states furnish a direct sum of representations. The same is true for the two sets of states 
$\{ |\beta_j^{(+)}(\vec{k}) \rangle\}$ and $\{ |\beta_j^{(-)}(\vec{k}) \rangle\}$. We have thus demonstrated the split of the photon discrete series representation into a direct sum of irreducible discrete series representations with opposite helicity, as shown in (\ref{discrete series in 4D split into two}).



\section{Conclusions and Outlook}
\label{sec:Conclusions and Outlook}

\paragraph{Summary.} In this work, we considered the free Maxwell gauge potential on $dS_4$. We fixed the gauge completely for the classical bulk gauge potential and then we quantized the propagating degrees of freedom by choosing the Bunch-Davies vacuum. We also explained how quantum dS charges act on single-photon states. Then, we took the late-time limit of our quantum field and we identified two late-time operators, $\alpha_j(\vec{k})$ and $\beta_j(\vec{k})$. We gave explicit expressions for the late-time operators in terms of bulk creation and annihilation operators, which we present here again for convenience,
\begin{subequations} 
    \begin{align}
 \alpha_j(\vec{k})&=i {{\frac{1}{\sqrt{2}}}}\sum_{\lambda = \pm }~\hat{\epsilon}_{j}^{(\lambda)}(\vec{k})~\left[\admk-a_{\vec{k},\lambda}\right]k^{-1/2}, ~~\Delta=1,\\
   \beta_j(\vec{k})&={{\frac{1}{\sqrt{2}}}}\sum_{\lambda =\pm} ~\hat{\epsilon}_{j}^{(\lambda)}(\vec{k})~\left[\admk +a_{\vec{k},\lambda}\right]k^{1/2},~~\Delta=2.
    \end{align}
\end{subequations}
In section \ref{sssec:The corresponding discrete series late-time operators}, we identified the $SO(4,1)$-invariant and positive-definite inner products with respect to which states of the form $ \alpha_j(\vec{k})|0\rangle$ and $\beta_j(\vec{k}) | 0 \rangle$ furnish the unitary discrete series series representation $D_{01}$ associated with the photon on $dS_4$ [see (\ref{isomorphisms}) and (\ref{discrete series in 4D split into two})]. We also explained how the inner product for $\beta_j$ can be `imported' from $\alpha_j$ by exploiting the isomorphisms (\ref{isomorphisms}).

Then, after introducing the self-dual and anti-self-dual parts of the photon field strength in section \ref{sec:field strength and anti-self-duality} (and explaining that these two parts do not mix with each other under dS transformations), we identified the gauge-potential late-time operators associated with the self-dual and anti-self-dual sectors in section \ref{sec:Direct sum of representations in the Hilbert space}. Let us give the corresponding late-time operators here again for convenience, 
\begin{subequations} \label{eqn:concl hel late time}
    \begin{align}
 \alpha_j^{(\lambda)}(\vec{k})&=i {{\frac{1}{\sqrt{2}}}}\left[\hat{\epsilon}_{j}^{(\lambda)}(\vec{k})\admk-\hat{\epsilon}_{j}^{(-\lambda)}(\vec{k})a_{\vec{k},-\lambda}\right]k^{-1/2}, ~~\Delta=1\\
   \beta_j^{(\lambda)}(\vec{k})&={{\frac{1}{\sqrt{2}}}}\left[\hat{\epsilon}_{j}^{(\lambda)}(\vec{k})\admk+\hat{\epsilon}_{j}^{(-\lambda)}(\vec{k})a_{\vec{k},-\lambda}\right]k^{1/2},~~\Delta=2.
    \end{align}
\end{subequations}
The $\lambda=+$ operators in  \eqref{eqn:concl hel late time} are the late-time gauge potentials for the anti-self-dual field strength, while the $\lambda=-$ ones are the late-time potentials for the self-dual field strength.

Last, we showed that the two sets of fixed-helicity states $\{\alpha_j^{(-)}(\vec{k}) |0 \rangle \}$ and $\{\alpha_j^{(+)}(\vec{k}) |0 \rangle \}$ furnish a direct sum of unitary discrete series irreducible representations of $SO(4,1)$ associated with photons of helicity $-1$ and $+1$, respectively. The same holds for the two sets of states $\{\beta_j^{(-)}(\vec{k}) |0 \rangle \}$ and $\{\beta_j^{(+)}(\vec{k}) |0 \rangle \}$. To sum up, for the normalized late-time operators we have,
\begin{subequations}\label{eqn:alphaNpm}
    \begin{align}
        discrete~series-:\alpha_j^{N(-)}(\vec{k})&=i\left[\hat{\epsilon}_{j}^{(-)}(\vec{k})a^\dagger_{-\vec{k},-}-\hat{\epsilon}_{j}^{(+)}(\vec{k})a_{\vec{k},+}\right]k^{-1/2},~~\Delta=1\\
        discrete~series+:\alpha_j^{N(+)}(\vec{k})&=i\left[\hat{\epsilon}_{j}^{(+)}(\vec{k})a^\dagger_{-\vec{k},+}-\hat{\epsilon}_{j}^{(-)}(\vec{k})a_{\vec{k},-}\right]k^{-1/2},~~\Delta=1.
    \end{align}
\end{subequations}
and similarly for $\beta_j^{N(\pm)}(\vec{k})$.


\paragraph{Commutation relations for late-time operators.} The two late-time operators, $\alpha_{j}$ and $\beta_j$, have non-trivial commutation relations. This is another feature that distinguishes the late-limit of dS field theories from the boundary limit of AdS field theories (in the AdS case, one of the two late-time operators of the Maxwell potential corresponds to a non-normalizable mode and is, therefore, ignored). The commutation relations of our fixed-helicity late-time operators are
\begin{subequations}\begin{align}
    \nonumber\left[\alpha^{N(\lambda)}_m(\vec{k}),\beta^{N(\lambda')}_l(\vec{k}')\right]&=-{i}(2\pi)^3\delta^{(3)}\left(\vec{k}+\vec{k}'\right)~\delta_{\lambda,-\lambda'}\\
    &\times \left( \delta_{ml}-\frac{k_m  k_l}{k^2} \right),\\
    \left[\alpha^{N(\lambda)}_m(\vec{k}),\alpha^{N(\lambda')}_l(\vec{k}')\right]&=(2\pi)^3~\delta^{(3)}\left(\vec{k}'+\vec{k}\right) ~\delta_{\lambda, - \lambda'}\\
    \nonumber&\times~i \lambda ~k^{-2}\,{k^a}~\epsilon_{mla}. \\
    \left[\beta^{N(\lambda)}_m(\vec{k}),\beta^{N(\lambda')}_l(\vec{k}')\right]&= (2\pi)^3~\delta^{(3)}\left(\vec{k}'+\vec{k}\right)~\delta_{\lambda, -\lambda'}\\
    \nonumber&\times~i \lambda ~\,{k^a}~\epsilon_{mla},
\end{align}
\end{subequations}
where we used $\hat{\epsilon}^{(\lambda)} (- \vec{k}) = \hat{\epsilon}^{(-\lambda)} ( \vec{k})$, as well as
\begin{align}
  &  -\hat{\epsilon}_l^{(-\lambda)}(\vec{k})\,\hat{\epsilon}_m^{(\lambda)}(\vec{k}) +\,\hat{\epsilon}_m^{(-\lambda)}(\vec{k}) \hat{\epsilon}_l^{(\lambda)}(\vec{k})\,\,=i \lambda \frac{k^a}{k}~\epsilon_{mla},\\
   &  \hat{\epsilon}_m^{(\lambda)}(\vec{k})\hat{\epsilon}_l^{(-\lambda)}(\vec{k}) +\, \hat{\epsilon}_l^{(\lambda)}(\vec{k})\,\hat{\epsilon}_m^{(-\lambda)}(\vec{k})\,=\delta_{ml}-\frac{k_m  k_l}{k^2} .
\end{align}
The first of these equations is a straightforward consequence of (\ref{eqn:epsilon cross prop}). The second equation coincides with (\ref{completeness of polarisation vecs}) because $\hat{\epsilon}_m^{(-\lambda)}(\vec{k}) = \hat{\epsilon}_m^{(\lambda)}(\vec{k})^*$.

  \paragraph{Other literature on the late-time realization of the dS photon representation.} Let us discuss the relation of our work to \cite{sun2021note}, where the late-time operators that arise from a massless vector on dS spacetime are also briefly addressed. Our work can be viewed as complementary to  the discussion of \cite{sun2021note}, while we also offer additional insight and details\footnote{See also \cite{Silva_note, Penedones:2023uqc, Taylor:2026ase}.}. In particular, \cite{sun2021note} considers general number of dimensions and works in the Lorenz gauge. Moreover in \cite{sun2021note}, although the scaling dimensions of the two late-time operators are recognized, explicit expressions in terms of Bunch-Davies creation and annihilation operators are not given, and hence, the normalizability of states that we consider in the present work is not demonstrated. Also, to the best of our knowledge, demonstrating the split of the photon discrete series representation of $SO(4,1)$ into a direct sum of irreducible representations at the level of late-time operators is presented in our present paper for the first time. %
Note that we use ``opposite'' notation from the one used in \cite{sun2021note} -- our $\Delta=1$ operator $\alpha_j$ is denoted as $\beta_j$ in \cite{sun2021note}, while our $\Delta=2$ operator $\beta_j$ is denoted as $\alpha_j$ in \cite{sun2021note}. %

 Unlike in our paper, the Hilbert space in \cite{sun2021note} is constructed using wavefunctions in position space, $\psi_{\bar{\Delta}}(\vec{x})$. To be specific, for a given late-time operator, $\mathcal{O}^{(\Delta)}_j(\vec{x})$,%
\footnote{In our notation, we have $\mathcal{O}^{(\Delta=1)}_j(\vec{x}) = \alpha_{j}(\vec{x})$ and $\mathcal{O}^{(\Delta=2)}_j(\vec{x}) = \beta_{j}(\vec{x})$.} 
states $|\Psi \rangle$ are constructed in \cite{sun2021note}  as
\begin{align}
    |\Psi \rangle = \int d^3x ~\psi^{j}_{\bar{\Delta}}(\vec{x})~\mathcal{O}^{(\Delta)}_{j}(\vec{x})| 0 \rangle,
\end{align}
where the scaling dimension assigned to the wavefunction $\psi^{j}_{\bar{\Delta}}(\vec{x})$  is $\bar{\Delta}=d-\Delta$. In other words, the representation of the dS group furnished by states created by the dimension-$\Delta$ operator, $\mathcal{O}^{(\Delta)}_{j}(\vec{x})| 0 \rangle$, can be alternatively studied in terms of wavefunctions $\psi^{j}_{\bar{\Delta}}(\vec{x})$.

Last, unlike \cite{sun2021note}, we first fixed the gauge completely in the bulk, and then we took the late-time limit of the bulk (fully gauge-fixed) quantum field. This is why our late-time states were found to furnish representations that were automatically in the quotient space realization. In particular, as shown in the main text, the states $| \alpha_j(\vec{k})\rangle$ furnish the photon unitary discrete series representation of $SO(4,1)$, with the CFT-inspired exceptional series inner product. This representation is labeled by $\Delta=2,s=1$ in \cite{sun2021note}, and is denoted as $\mathcal{U}_{1,0} = \mathcal{F}_{1,1}/Im(z \cdot \partial_x)$ -- this corresponds to $C^{'-}_{01}/F'_{01}$ in our notation (\ref{isomorphisms}).


\subsection{Possible relation to boundary conformal gauge fields in the microscopic realization of dS higher-spin gravity} \label{subsec: boundary gauge fields and currents}

The late-time asymptotic expansion of the Maxwell field  gives rise to two boundary operators, $\alpha_j$ (leading) and $\beta_j$ (subleading), with conformal dimensions 1 and 2, respectively. Following the standard AdS terminology, one would identify $\alpha_j$
 with a boundary gauge field and $\beta_j$ with the conserved current. However, unlike in anti-de Sitter, where the leading mode is non-normalizable and is therefore regarded as fixed boundary data, in dS both asymptotic modes are a priori part of the late-time data. Our construction employs dS-invariant late-time inner products, with respect to which the Hilbert space is furnished  by states created by $\alpha_j$, as well as by $\beta_j$. Moreover, the two dS late-time operators have non-trivial commutation relations, as shown earlier.

In what follows, let us briefly discuss an exotic theory of quantum gravity in dS space where our spin-1 operators are expected to play a role. In particular, in the context of microscopic realizations of dS space, a convenient (and concrete) model is higher-spin gravity on $dS_4$ \cite{Anninos:2011ui, Anninos:2017eib, Anninos:2026hia, Neiman:2014npa, David:2020ptn, Neiman:2018ufb, Vasiliev:1990en, Anninos:2019nib, De:2026shn}. The so-called non-minimal  theory of higher-spin gravity includes an infinite tower of integer-spin massless fields in its spectrum, with spins: $s=0$ (conformally coupled scalar), $s=1$ ($\sim$ photon), $s=2$ ($\sim$ graviton) and so forth. This theory is consistent around a $dS_4$ background and its microscopic/quantum completion has been discussed in \cite{Anninos:2026hia, Anninos:2017eib}. In particular, from \cite{Anninos:2026hia} it follows that (in the Euclidean path integral approach) the microscopic theory underlying the non-minimal higher-spin gravity on $S^4$ corresponds to the Euclidean path integral of the $U(N)|_{N \to -N}$ model\footnote{The free $U(N)$ model is a theory of $N$ free complex conformally coupled scalar fields. By $U(-N)$ model we mean that the statistics of the scalar fields is reversed.} on $S^3$ coupled to higher-spin sources ($\sim$ conformal gauge fields). In particular, the proposal of \cite{Anninos:2026hia}, verified up to 1 loop, is that there are two such microscopic $S^3$ path integrals `glued' together through path integration over the conformal gauge fields.

Leaving the path integral picture of \cite{Anninos:2026hia} aside, we can pose the question: how can we reproduce the features of the Euclidean microscopic realization from the perspective of  Lorentzian planar $dS_4$?%
\footnote{For recent interesting findings concerning higher-spin in the cosmological (dS) context, see \cite{De:2026shn}.} %
In order to achieve this, let us recall that, in the context of Lorentzian planar $dS_4$, the microscopic theory that is proposed to replace the $U(N)|_{N \to -N}$ model lives on the late-time boundary, and is known as the $Q$-model \cite{Anninos:2017eib}%
\footnote{The $Q$-model of \cite{Anninos:2017eib} refers to the minimal version of higher-spin gravity. Here, for the sake of our discussion, we refer to the $Q$-model appropriately generalized to the case of non-minimal higher-spin gravity.}. %
The $Q$-model includes $N$ complex microscopic operators $Q^{(i)}(\vec{x})$ ($i=1,...,N$) that transform as conformal primaries under $SO(4,1)$ with scaling dimension $\Delta=1/2$, and their `conjugate momenta' $P^{(i)}(\vec{x})$. Thus, a concrete question one can pose is:
\\
\\
\textit{Can we uncover the existence of conformal gauge fields (which play the role of the `glue' in \cite{Anninos:2026hia}) in the $Q$-model of \cite{Anninos:2017eib}?}
\\
\\
Although we do not give the answer here, let us discuss some possible directions. First, note that in \cite{Anninos:2017eib}, the late-time profiles of the bulk fields of higher-spin gravity correspond to bilinears constructed from the $Q$-fields and their conjugate momenta.  For example, in the massless spin-1 case which is relevant to our present paper, we expect that the subleading, $\Delta=2$, operator ($\sim$ conserved current) is related to the $Q$-fields schematically as \cite{Anninos:2017eib, Anninos:2019nib}
\begin{align}\label{Q-model U(1) current}
  \beta_j(\vec{x}) \sim \sum_{a=1}^{N}\left(Q^{(a)\dagger}(\vec{x})~\partial_jP^{(a)}(\vec{x})  -\partial_j Q^{(a)\dagger}(\vec{x})~P^{(a)}(\vec{x})  ~\right)  + (h.c.) .
\end{align}
It is important to note that the $Q$-model comes equipped with its own microscopic norm and Hilbert space structure. Using the microscopic `$Q$-model norm',  we expect that the spin-1 operator on the right-hand side of (\ref{Q-model U(1) current}) furnishes the unitary discrete series representation of $SO(4,1)$ associated with the photon. An expression for the leading, $\Delta=1$, operator $\alpha_j$ similar to (\ref{Q-model U(1) current}) can be also found \cite{Anninos:2017eib}. What we would like to address in future work is to investigate whether, by using the $Q$-model norm (which, in principle, does not coincide with the CFT-inspired norm we used for dS late-time operators), states created with {$\alpha_j$} furnish the representation of a conformal gauge field in 3 dimensions which carries no propagating degrees of freedom (and, thus, the corresponding states are expected to have zero norm). Some further discussions concerning conformal gauge fields in 3 dimensions are given below.

 Before concluding, let us discuss how a non-local Lagrangian for a self-interacting  spin-1 conformal gauge  field $\mathcal{A}_j$ in 3 dimensions can be found \cite{Giombi:2013yva}.%
 \footnote{Related discussions on spinning conformal gauge fields can be found in, e.g., \cite{Anninos:2026hia, Anninos:2020hfj, Beccaria:2014jxa, Giombi:2013yva}.} %
 One considers, for example, a Euclidean CFT of $N$ complex massless scalar fields $\phi^{(a)}$ ($a=1,...,N$) on $\mathbb{R}^3$ coupled to a real background $U(1)$ gauge field:
 \begin{align}
     S= \int d^3x \sum_{a=1}^{N}\Big( \partial_j\phi^{(a)*}~\partial^{j}\phi^{(a)} +  i\left(\phi^{(a)*}\partial_j\phi^{(a)}  -\phi^{(a)}  ~\partial_j\phi^{(a)*}\right)\mathcal{A}^j ~ {+\mathcal{A}_j\,\mathcal{A}^j~\phi^{(a)*}\phi^{(a)}\Big)}.
 \end{align}
 {The action is invariant under the simultaneous transformations $\mathcal{A}_j(\vec{x}) \to \mathcal{A}_j(\vec{x}) 
 + \partial_{j}c(\vec{x})$  and $\phi^{(a)}(\vec{x}) \to e^{i\,c(\vec{x})} \phi^{(a)}(\vec{x})$.}
 Integrating out the scalar fields, we find an induced  non-local action for the conformal gauge field which is invariant under the gauge transformation $\mathcal{A}_j(\vec{x}) \to \mathcal{A}_j(\vec{x}) 
 + \partial_{j}c(\vec{x})$. The theory admits a perturbative description for large $N$. The action for the linearized fluctuation $\delta \mathcal{A}_j$ describes a free conformal spin-1 gauge field with no local propagating degrees of freedom, while the $SO(4,1)$ representation associated with this might be identified with the representation (realized in terms of zero-norm states) furnished by our late-time operator $\alpha_j$ using the appropriate microscopically defined norm (as in the case of the $Q$-model). The fact that the linearized spin-1 conformal gauge field carries no propagating degrees of freedom in 3 dimensions stems from the fact that its field equations have only longitudinal solutions and such solutions are pure gauge \cite{Giombi:2013yva}.


\section*{Acknowledgment}
We thank Dionysios Anninos, Benjamin Pethybridge, Guillermo Silva, Zimo Sun and Joseph Lap  for insightful discussions. We also thank the anonymous referee for their comments.
The work of V. A. L. was supported by the ULYSSE Incentive Grant for Mobility in
Scientific Research [MISU] F.6003.24, F.R.S.-FNRS, Belgium. M.B., E. B. G. and G.Ş. acknowledge support from TÜBİTAK 2232 - B International Fellowship for Early Stage Researchers program with
project number 121C138. G.Ş. also acknowledges support from Boğaziçi University Start Up Grant no
19858 in the final stages of this work and support by the Munich Institute for Astro-, Particle and BioPhysics (MIAPbP) which is funded by the Deutsche Forschungsgemeinschaft (DFG, German Research Foundation) under Germany´s Excellence Strategy – EXC-2094 – 390783311 during the Primordial Cosmology program. V. A. L. and G. Ş. also thank the workshop ``Quantum de Sitter Universe" (DAMTP), funded by the Gravity Theory Trust and the Centre for Theoretical Cosmology, which offered a nice environment for inspiring discussions.\\
\\

\appendix

\providecommand{\href}[2]{#2}\begingroup\raggedright\endgroup

\end{document}